\documentclass{article}

\usepackage{arxiv}

\usepackage[utf8]{inputenc} % allow utf-8 input
\usepackage[T1]{fontenc}    % use 8-bit T1 fonts
\usepackage{hyperref}       % hyperlinks
\usepackage{url}            % simple URL typesetting
\usepackage{booktabs}       % professional-quality tables
\usepackage{amsfonts}       % blackboard math symbols
\usepackage{nicefrac}       % compact symbols for 1/2, etc.
\usepackage{microtype}      % microtypography
\usepackage{lipsum}		% Can be removed after putting your text content
\usepackage{graphicx}
\usepackage{doi}

\usepackage{amsmath}
\usepackage{amssymb}
\usepackage{mathrsfs}
\usepackage{dsfont}
\usepackage{framed}

\newcommand{\ltfrac}[2]{\mbox{\large$\frac{#1}{#2}$}}

\usepackage{graphicx}
\graphicspath{ {./images/} }

\usepackage{graphicx}
\usepackage[table]{xcolor}

\usepackage{url}
\usepackage{cleveref}
\usepackage{hyperref,xcolor}
\hypersetup{
colorlinks,
linkcolor={red!50!black},
citecolor={blue!50!black},
urlcolor={blue!70!black}
}

\title{Similarity Suppresses Cyclicity: \\ Why Similar Competitors Form Hierarchies}

\author{Christopher Cebra \\
	Department of Statistics\\
	University of Chicago\\
	Chicago, IL 60637 \\
	 \\
	\And
	\href{https://orcid.org/0000-0001-6618-631X}{\includegraphics[scale=0.06]{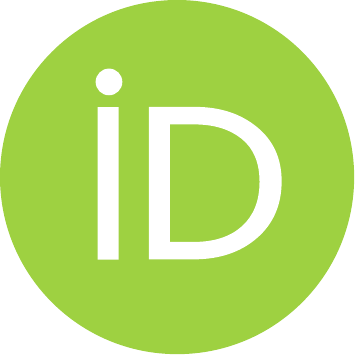}\hspace{1mm}Alexander Strang} \\
	Department of Statistics\\
	University of Chicago\\
	Chicago, IL 60637 \\
	\texttt{alexstrang@uchicago.edu} \\
}

\renewcommand{\shorttitle}{ Similarity Suppresses Cyclicity}

\hypersetup{
pdftitle={Similarity Suppresses Cyclicity},
pdfsubject={q-bio.NC, q-bio.PE},
pdfauthor={Christopher Cebra, Alexander Strang},
pdfkeywords={Competitive Systems, Evolutionary Game Theory, Transitivity, Cyclic Competition, Helmholtz-Hodge Decomposition},
}

\begin{document}
\maketitle

\begin{abstract}
	Competitive systems can exhibit both hierarchical (transitive) and cyclic (intransitive) structures. Despite theoretical interest in cyclic competition, which offers richer dynamics, and occupies a larger subset of the space of possible competitive systems, most real-world systems are predominantly transitive.  Why?  Here, we introduce a generic mechanism which promotes transitivity, even when there is ample room for cyclicity. Consider a competitive system where outcomes are mediated by competitor attributes via a performance function. We demonstrate that, if competitive outcomes depend smoothly on competitor attributes, then similar competitors compete transitively. We quantify the rate of convergence to transitivity given the similarity of the competitors and the smoothness of the performance function. Thus, we prove the adage regarding apples and oranges. Similar objects admit well ordered comparisons. Diverse objects may not. To test that theory, we run a series of evolution experiments designed to mimic genetic training algorithms. We consider a series of canonical bimatrix games and an ensemble of random performance functions that demonstrate the generality of our mechanism, even when faced with highly cyclic games. We vary the training parameters controlling the evolution process, and the shape parameters controlling the performance function, to evaluate the robustness of our results. These experiments illustrate that, if competitors evolve to optimize performance, then their traits may converge, leading to transitivity.
\end{abstract}

% keywords can be removed

%%%%%%%%%%%%%%%%%%%%%%%%%%%%%%%%%%%%%%%%%%%%%%%%%%
%% main text

\section{Introduction} 

\subsection{Motivation } \label{sec: motivation}

The space of possible competitive systems contains significantly more cyclic systems than are observed in practice. Why?

% Most real world competitive systems are predominantly transitive. Most possible systems are predominantly cyclic. Why?

%Why are most competitive systems predominantly transitive?

%% introduce competitive systems
A competitive system is a collection of agents who compete with one another. Competitive systems abound, including examples in biology, politics, economics, sports, and artificial intelligence. Their structures are encoded in the collection of advantages possessed by certain competitors over others. If events occur pairwise, then the structure of advantages and disadvantages can be represented with a network. Agents are assigned to nodes, edges connect agents who compete, and edges are weighted by values representing advantage. 

%% discuss the distinction between transitive and intransitive systems
Competitive systems are transitive if the network of advantage is consistent with some rank order from best to worst. Then competitors can be arranged into a hierarchy. Hierarchy and transitivity are linked by a consistency condition. Suppose that, if $A$ possesses an advantage over $B$, and $B$ possesses an advantage over $C$, then $A$ possesses and advantage over $C$. Then advantage is transitive, and we can order the agents. If, instead, $C$ possesses an advantage over $A$, then the competitive system is cyclic, and the competitors form a rock-paper-scissor type cycle. In this case, competition is intransitive, and the competitors cannot be consistently ranked. Competition is transitive if and only if there is no such cycle in the network.

%% connect to rating and a notion of quality/fitness. natural model for competition 
Competitive systems are largely understood via ranking. In sports, rankings are widely published, dictate draft orders and post-season schedules, and are eagerly consumed by fans. 
Outside of sports, ranking plays an important role in decision problems (e.~g.~pairwise choice, social choice) and data science. 
Examples in data science include ranking colleges \cite{Luca,Mcdonough,Monks}, web search results \cite{Brin, Bryan}, and  movie suggestions on streaming platforms \cite{Bell_c,Bell_a,Bell_b}. 
Examples in decision making include elections where results are determined by rank.
Examples in biology include hierarchical animal societies, where dominance is often associated with priority access to resources \cite{Drews,Klass_b,Koenig,Solberg}, territory maintenance \cite{Shoemaker}, and higher reproductive output \cite{Haley,Muniz}.
Ranking also plays a key role when training artificial intelligences, who must rank scenarios in order to make choices, and who are themselves often ranked during the training process.
Thus, ranking plays an essential role in our understanding of competitive systems across domains.

When competitors can be ranked, we can unambiguously answer which are better or worse. Then there may implicitly exists some notion of inherent competitor competence which is compared to determine advantage. In other words, it is meaningful to ask, how good is agent $A$ at the game? Familiar notions, like competitor fitness, often presume the existence of such an opponent independent measure of competence. Ratings assign each agent a measure of competence, while rankings order the agents. Ranking is often performed by first rating the competitors, then listing them from highest to lowest. Such methods are widely studied, and vary depending on the field of interest (c.f.~\cite{Bozoki,Brin,Keener,Kwiesielewicz_a,Kwiesielewicz_b, Langville,Massey,Stefani_c,Stefani_a,Stefani_b}). Nevertheless, the underlying conceptual model remains unchanged. Competitive advantage is determined by a comparison of competitor quality.

The transitive model is so pervasive that systems that cannot be consistently ranked are almost universally treated as surprising or disturbing. In psychology, economics, and social choice theory, cyclic preferences in opinion are considered ``irrational'', ``paradoxical'', and ``chaotic'' \cite{Mckelvey_a,Mckelvey_b,Regenwetter_d}.  In biology, cycles are treated with less alarm, but with equal surprise and interest. Theoretical work suggests that cyclic competition may maintain biodiversity by preventing competitive exclusion \cite{Laird,May,Reichenbach_a,Reichenbach_b,Reichenbach_c,Reichenbach_d,Xue}, and may lead to deeply counter-intuitive evolutionary dynamics such as ``survival of the weakest'' \cite{Frean}. %In some situations, apparent intransitivity may arise incidentally, as a result of measurement error or randomness in event outcomes. \cc{I feel like this section could use an example, which I see you had footnoted, although that may need to be cleaned up.} %\footnote{For example, in the 2019 season of Major League Baseball, the Houston Astros beat the Seattle Mariners (18 of 19 games) who beat the Pittsburgh Pirates (3 of 3 games) who in turn beat the Astros (2 of 3 games). Given the small event counts and close game margins it this cycle is easily explained by chance.}
Popular empirical examples include Sinervo's side-blotched lizards \cite{sinervo1996rock}, or Kerr's colicin producing E. Coli \cite{Kerr}. When inherent, cycles can alter long term dynamics \cite{Reichenbach_a,Reichenbach_b,Reichenbach_c,Reichenbach_d}, and optimal strategies \cite{candogan2011flows}. In a decision making context such as an election, the extent of cycles will determine to what extent the outcome of elections can be influenced by strategic deal-making \cite{Lagerspetz}, the order in which choices are presented \cite{Flanagan}, or the individuals with agenda setting authority \cite{Morse}.

The vast majority of possible competitive systems are intransitive, exhibit cycles, and cannot be consistently ranked or rated without admitting errors. From this perspective the transitive lens is woefully insufficient. It comes far short of describing all possible advantaged networks. Nevertheless, most real world competitive systems are predominantly transitive, with little to no evidence of statistically significant cycles. For example, when Go is played among a diverse ensemble of agents trained to explore the policy space it is highly cyclic, yet real Go playing agents trained by Deepmind \cite{silver2016mastering} compete transitively \cite{omidshafiei2020navigating}. Thus, while a transitive lens may appear narrow in theory, it often succeeds in practice. 

The startling paucity of cycles in the face of their overwhelming possibility has driven longstanding debates. Theorists, compulsed by counterexamples, often emphasize the complexities that cycles can produce, while empiricists argue that cycles, and their ensuing complexities, are rarely of great concern. %One camp, typically empirically motivated, argues that most tournaments are amenable to ranking and do not exhibit cyclic behavior. The other, supported by theory and select case-studies, argues that  not all tournaments can be ranked, cycles play an important role in some systems, and that the ranking perspective is an impoverished simplification of real systems. 

As illustration, consider social choice theory and elections. Cycles in opinion play an important role in social choice theory. When aggregate voter opinions involving the top candidates are cyclic, no election system can guarantee that the winner fairly reflects the majority opinion \cite{Arrow,van_Deemen_b}. Such cycles can occur even if each voter's preferences are well ordered.
% explain Condorcet paradox: occurs b/c of top cycle, aggregate preferences produce cycles
This situation is an example of Condorcet's paradox.\footnote{A Condorcet paradox occurs when there is no Condorcet winner - a candidate who would defeat any other candidate in a head-to-head election \cite{Gehrlein_c}. A simple example suffices. Suppose that one voter prefers $A$ to $B$ to $C$, the next prefers $B$ to $C$ to $A$, and the third prefers $C$ to $A$ to $B$. Then $A$ would beat $B$ in a head-to-head election, who would beat $C$, who would beat $A$.} %Extensive theoretical literature has examined the likelihood of voting cycles, necessary conditions which forbid cycles, and the implications of cycles on the impossibility of fair election procedures \cite{Arrow,Black,Gehrlein_d,Gibbard,Reny,Satterthwaite,Sen,van_Deemen_b}.
 Voter cycles have been observed in a number of historical case studies \cite{Gehrlein_c} including voting on the annexation of Texas \cite{Morse}, the subsequent status (free or slave) of land gained after the Mexican-American war \cite{Gehrlein_c,Riker_a}, abortion reform in Canada \cite{Flanagan}, and public opinion on U.S.~intervention in Kuwait preceding the Gulf War  \cite{Gaubatz}. Axiomatic social choice theory emphasizes the ``impossibility" of aggregating voter opinion given the large space of preferences that cannot be fairly aggregated. If voter opinion is uniformly distributed, then the chance of a voter cycles increases in the number of voters and candidates \cite{Gehrlein_d}.

% relevance of Condorcet is debated
Despite these examples, the empirical relevance of Condorcet's paradox is controversial \cite{Kurrild_a,Kurrild_c,Munkoe,van_Deemen_b}.
% state that previous examples are anecdotal at best, and often rely on inferred voter preferences
Case-studies only provide anecdotal evidence and often rely on reconstructed voter preferences (c.f.~\cite{Gaubatz}).
% studies large electorates
Multiple authors have evaluated the frequency of voting cycles in large electorates using empirical preference data \cite{Kurrild_a,Munkoe,Tideman,van_Deemen_a,van_Deemen_b}. 
% van Deemen and Gehrlein meta-studies, rare but occasionally seen in elections, rarer in large electorates, note distinction in top cycle vs transitivity in general
Combined, these studies indicate that Condorcet's paradox rarely occurs in large electorates, with most studies finding few if any cycles. In a meta-analysis, Van Deemen \cite{van_Deemen_a} and Gehrlein \cite{Gehrlein_b} found that roughly ten percent of 265 elections studied exhibited the paradox \cite{van_Deemen_b}, usually among small electorates. Thus, the axiomatic theory been criticized for overstating the prevalence of voter cycles \cite{Regenwetter_a,Regenwetter_b}.

%% this same story repeats in other fields
The same story repeats across fields (c.f.~\cite{Laird,Soliveres} vs.~\cite{Godoy}, or \cite{chen2016modeling}). Most real competitive systems are mostly transitive, even though most possible competitive systems are mostly cyclic. Which begs the question; \textbf{\textit{why}?} What mechanisms produce transitive systems without assuming transitivity a priori?

%% discuss domain specific mechanisms (social choice)
Domain specific mechanisms are well studied. In social choice theory, certain domain restrictions on voter preferences guarantee transitivity \cite{Black,Sen}. If the choices in an election can be arranged on a single axis, and all voter preferences are single peaked, then the aggregate preferences are necessarily transitive and the Condorcet winner is the median voter's favorite \cite{Black}. These domain restrictions are frequently violated in empirical studies so do not constitute a plausible explanation for the infrequency of cycles \cite{Kurrild_c,Regenwetter_b}. An alternate body of theory considers the probability of observing cycles given a particular ``culture", i.e.~a distribution of voter preferences.  In an impartial cultures, voter preferences are highly heterogeneous, preventing easy aggregation. Under more realistic assumptions, voter preferences are more homogeneous and mutually correlated, easing aggregation and reducing the chance of observing cycles \cite{Regenwetter_a}.

In biology, social hierarchies are wide-spread, and are sustained by diverse mechanisms. While dominance in social hierarchies is highly advantageous, escalation in agonistic interactions is often dangerous. Thus, psychology and social conventions play a large role in maintaining and promoting hierarchies \cite{Landau_a,Landua_b}. Individuals size up their opponents based on past events. Opponent evaluation leads to winner, loser, and bystander effects, where the outcome of past events informs behavior during future events. These effects can establish self-organizing hierarchies, even with highly random initial outcomes, or nearly equivalent individuals \cite{Chase_b,Chase_a,Hsu,Oliveira,Silk}. In some extreme cases the costs of escalation to the loser, or to a society, drive populations to adopt conflict resolution conventions, even if the convention is removed from individual fitness. For example, hyena societies adopt matrilineal rank inheritance \cite{Smale,Strauss}.

\begin{figure}[t!]
		\centering
		\includegraphics[scale=0.3]{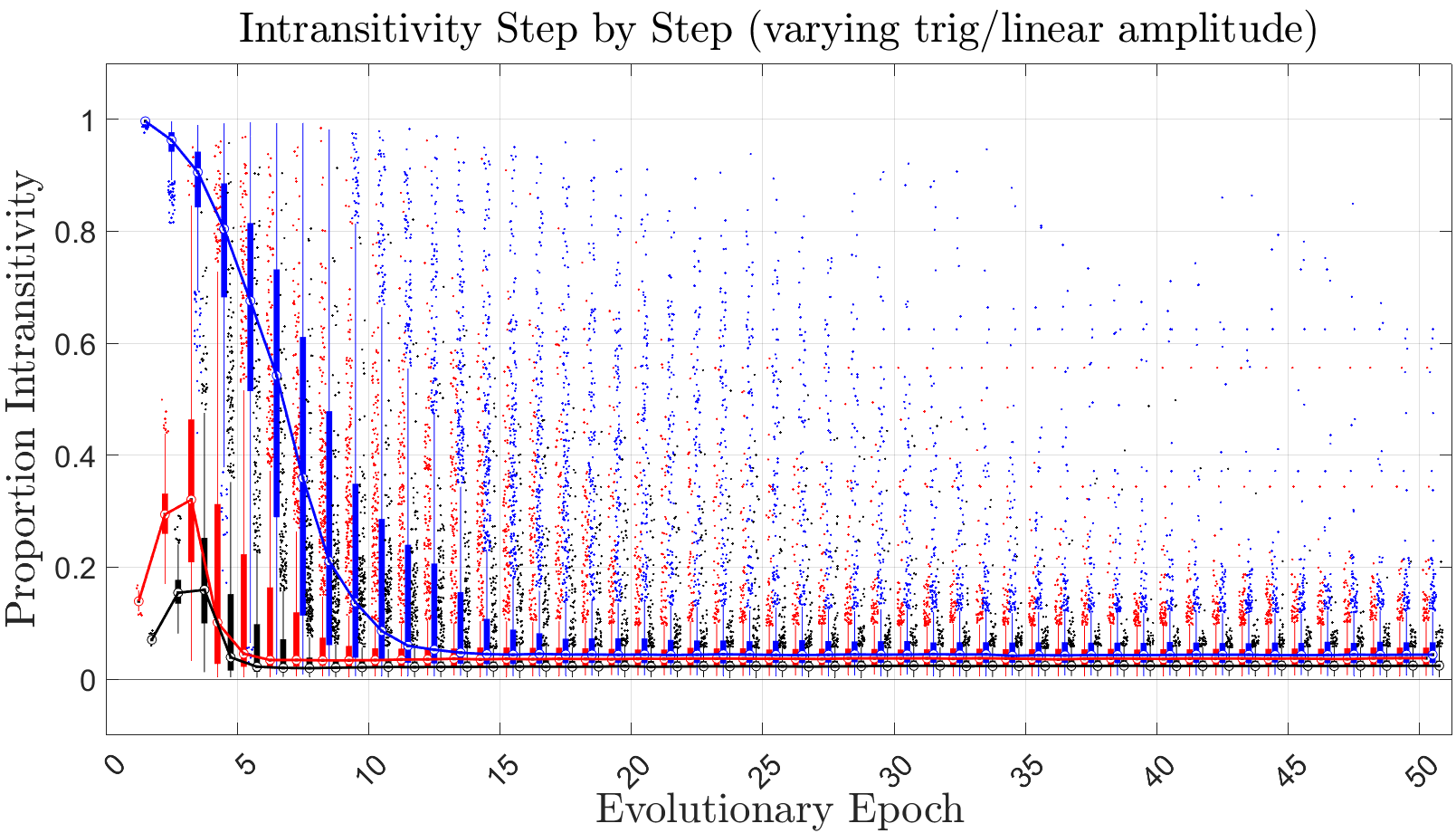}
		\caption{Decaying intransitivity across evolution for an ensemble of random performance function. The uppermost blue line represents a purely trigonometric performance function, which admits cycles for diverse populations. The intermediate red line represents a mixture between a trigonometric and a linear performance function, and the lower black line represents a close to linear performance function. In general, the closer to trigonometric, the rougher the function, thus the more cycles it admits. For further information about the structure of performance functions and how we selected them, see Section \ref{sec: numerics}. }
		\label{fig: linear and trig amplitude intransitivity}
\end{figure}

These mechanisms are highly domain dependent. Nevertheless, transitivity is observed in simpler settings where the space of competitive systems ought to make transitivity unlikely. For example, Czarnecki et al.~tracked competitive structure over the course of training a population of a.i.~agents in a series of games ranging from Connect Four and Tic Tac Toe, to Go and Starcraft. They observed that most games exhibit a ``spinning top structure", wherein the bulk of policies are of intermediate ability, a small proportion achieve high performance, cyclic structures dominate among mediocre agents, but excellent agents compete transitively. Thus, training proceeds through a cyclic intermediate stage composed of a diverse population, before arriving at a rarefied collection of exceptional agents who compete transitively \cite{czarnecki2020real}. Convergence towards transitive populations occurs without any external hierarchy promoting mechanisms, and is instead a feature of the underlying games. Czarnecki et al.~suggest that this structure is needed to make games interesting, challenging, and rewarding, thus is selected for. Such game structures have interesting consequences for training in multiplayer games (c.f.~\cite{omidshafiei2020navigating}).

%% we aim for a more general mechanism
Here we present a generic mechanism which promotes transitive competition, and suppresses cyclic competition, independent of any domain specific assumptions or hierarchy promoting conventions. It applies even when the space of possible systems is predominantly cyclic. Namely, if the relation between competitor advantage depends smoothly on competitor attributes, then sufficiently similar competitors compete transitively. Thus, similarity alone is enough to explain transitivity when advantage depends smoothly on attributes. More succinctly, \textbf{\textit{similarity suppresses cyclicity.}}

%We develop the necessary theory in Section \ref{sec: results}, then demonstrate the mechanism in a series of numerical experiments in Section \ref{sec: numerics}.
Figure \ref{fig: linear and trig amplitude intransitivity} shows a striking example of this concentration mechanism at work. We plot the observed intransitivity of a population of competitors over time as they evolve, for games of varying smoothness. Simulation details are provided in Section \ref{sec: random performance functions}. Over time, the populations concentrate, and the competitive networks approach perfect transitivity, even if the initial population is almost perfectly intransitive.

\subsection{Definitions} \label{sec: definitions}

\begin{figure}[t!]
		\centering
		\includegraphics[scale=0.45]{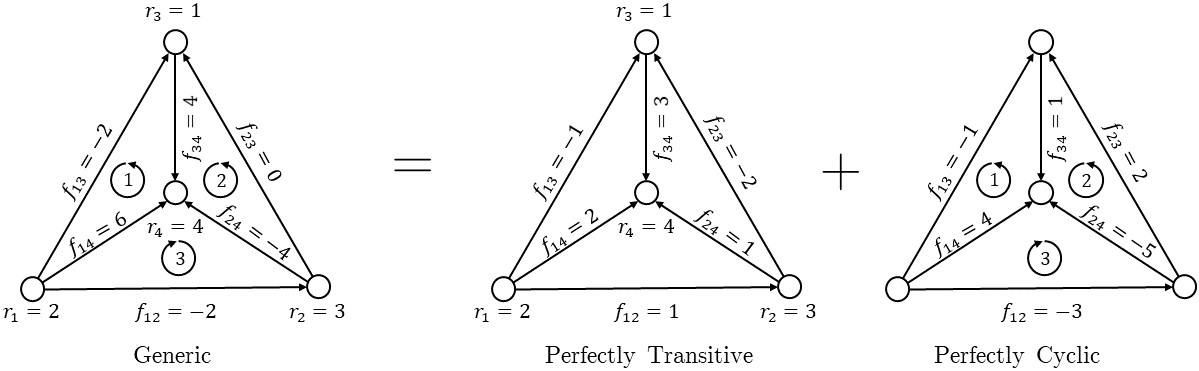}
		\caption{A visualization of the Helmholtz-Hodge decompostion applied to a complete, 4-node graph. Competitor ratings are expressed in the form $r_i$ for competitor $i$, while cyclic flow is denoted by the circular arrows, in the direction of and with the magnitudes listed.}
		\label{fig: HHD pedagogy}
\end{figure}

To study transitivity, we need to a metric to measure it by. We use the Hodge transitivity and intransitivity measures first proposed in \cite{jiang2011statistical} and further developed in \cite{strang2022network}. These metrics are defined in terms of a decomposition, namely, the discrete Helmholtz Hodge decomposition (HHD) \cite{jiang2011statistical, lim2020hodge,strang2022network}. The decomposition and associated measures are defined below. For details, see \cite{strang2022network}. A graphic of the Helmholtz-Hodge decomposition applied to a simplified competitive network can be found in Figure \ref{fig: HHD pedagogy}.

Consider an ensemble of agents who engage in pairwise events. Let $\mathcal{G} = (\mathcal{V},\mathcal{E})$ be a competitive network, with nodes $\mathcal{V}$ representing the agents. The network
$\mathcal{G}$ is undirected, with edges $\mathcal{E}$ connecting agents who compete. If all agents could compete, then the graph is complete. Competitive advantage is represented by adding an edge flow to the network. An edge flow is an alternating function on the edges, $f$, where $f_{ij}$ is the advantage agent $i$ possesses over agent $j$ \cite{jiang2011statistical,strang2022network}. The function $f$ is alternating if $f_{ij} = - f_{ji}$. Advantage can be measured using the difference in expected payout to agents $i$ and $j$, the log odds $i$ beats $j$, or some other monotonically increasing, skew symmetric function of the probability $i$ beats $j$. 

The HHD decomposes $f$ into two components, a transitive component $f_t$ associated with hierarchical competition, and a cyclic component $f_c$ that introduces intransitivity. Figure \ref{fig: HHD pedagogy} illustrates an example. The transitive component is consistent with a ranking and a rating. The transitive component on edge $ij$ equals the difference in a least squares rating $r$ of the agents $i$ and $j$. Then $f_t$ is curl free (does not cycle), and allows a quantitative measure of the quality of each agent via the ratings $r$. The rating of agent $i$ equals their average advantage over their neighbors, plus the average rating of their neighbors. Thus the ratings automatically account for neighborhood strength. If the $f = f_t$ and the edge flow is defined as the log odds of victory, then the ratings $r$ are Elo ratings, competition satisfies the Bradley-Terry model. When a edge flow can be expressed as the difference in some ratings assigned to the agents we say that the edge flow is perfectly transitive \cite{strang2022network}.

In contrast, $f_c$ encodes cyclic relations, like rock-paper-scissors triangles. It is favorite free, meaning that, if $f = f_c$, then no agent posses an average advantage or disadvantage relative to their neighbors. It is also cyclic in the sense that, any path moving along $f_c$ can be extended to close a loop without moving against $f_c$. We say a network is perfectly cyclic if it is both cyclic and favorite free. Then $f = f_c$. 

The spaces of perfectly cyclic and perfectly transitive edge flows are orthogonal complements that span the space of all possible edge flows. Thus, any competitive network admits a unique decomposition $f = f_t + f_c$ \cite{strang2022network}. The components $f_t$ and $f_c$ are each projections onto the perfectly transitive and cyclic subspaces. It follows that each is the best approximation to $f$ restricted to perfect transitivity or cyclicity. Then $\|f_c\|$ measures the distance from $f$ to its nearest perfectly transitive approximation, and $\|f_t\|$ measures the distance from $f$ to its nearest perfectly cyclic approximation. These are, respectively, the cyclicity and transitivity of the edge flow. As the cyclic component is responsible for introducing intransitivity, $\|f_c\|$ is sometimes called the intransitivity, or Hodge intransitivity. It is an absolute measure of the strength of the cyclic component of competition. The larger $\|f_c\|$ the more cyclic competition. 

The Hodge intransitivity is analogous to standard intransitivity measures, like the Slater measure which evaluates the distance (in edge reversals) to the nearest transitive network \cite{Slater}, or the Kendall measure \cite{Kendall}. Both the Kendall and Hodge measures can be derived from the variance across agents of the net advantage an agent possesses over their neighbors, provided the graph is complete. Unlike the Slater measure, the Hodge measure is easy to compute since the HHD can be performed by solving a sparse least squares problem, while the Slater measure requires solving an NP hard problem. Unlike the Kendall measure, the HHD is not limited to complete graphs, so is better adapted to empirical settings \cite{strang2022network}. 

The measures can be compared since $\|f_t\|^2 + \|f_c\|^2 = \|f \|^2$. Then $(\|f_t\|/\|f\|)^2$ and $(\|f_c\|/\|f\|)^2$ represent the proportions of competition that are transitive and cyclic. The relative measures are nonnegative, range from 0 to 1, and add to 1. %The relative measure $(\|f_c\|/\|f\|)^2$ is a more appropriate measure of intransitivity than $\|f_c\|$ since $\|f_c\|$ measures the size of a cyclic component of competition, which is entirely independent of $\|f_t\|$. Whether or not $f_c$ introduces significant intransitivity depends on whether $f_c$ is large relative to $f_t$. Increasing $f_t$ will always promote transitivity even if $f_c$, and thus $\|f_c\|$, is unchanged. In contrast, the relative measure $(\|f_c\|/\|f\|)^2$ will decrease whenever the transitive component grows, or increase when the cyclic component grows. The two relative measures represent a tradeoff between two opposite modes of competition and characterize which mode is better represented in the overall flow. Thus, $\|f_c\|$ should be considered a measure of cyclicity, while $(\|f_c\|/\|f\|)^2$ is measures of intransitivity. 
We use both the absolute and relative measures to quantify the overall structure of competitive advantage in a network. We use $(\|f_c\|/\|f\|)^2$ to quantify intransitivity, since transitivity is only violated if $f_c$ is large relative to $f_t$, thus makes up an appreciable portion of $f$. 

We also need language to describe the relation between agent attributes and agent performance. Suppose that competitive advantage is mediated by competitor attributes. Specifically, assume that the advantage $i$ possesses over $j$ depends on the attributes of $i$ and $j$, and some other outside factors that are shared on all agent pairs. Then competitive advantage can be expressed as a function of the attributes of $i$ and $j$ by marginalizing over any external influences. 

Let $\Omega \in \mathbb{R}^T$ be the region of admissible attributes in a $T$ dimensional trait space. Let $x,y%,z
$ represent trait vectors. Let $f$ be a performance function mapping from $\Omega \times \Omega \rightarrow \mathbb{R}$ where $f(x,y)$ is the advantage a competitor with traits $x$ possesses over a competitor with traits $y$. Then $f$ defines a functional form game \cite{balduzzi2019open,balduzzi2018re}. Functional form games model a wide variety of competitive systems in biology, social sciences, and artificial intelligence. Fairness requires that $f(x,y) = -f(y,x)$, otherwise input order confers arbitrary advantage. Applying the performance function to each edge returns the edge flow representing advantage.

In a functional form game the structure of advantage is determined by  who is playing. Any observed structure depends both on the function relating attributes to performance, and on the underlying distribution of players. A trait-performance model extends a functional form game by introducing a trait distribution $\pi_x$. If competitor attributes are sampled independently and identically from $\pi_x$, and advantage is determined by a performance function $f$, then the trait-performance theorem \cite{strang2022network} offers a set of simple statistical relations that predict the expected advantage structure. These relations are easy to interpret, and are explored in depth in \cite{strang2022network}. The theorem is reviewed in Section \ref{sec: trait performance}. We use those relations to show that similarity leads to transitivity when $f$ is smooth. 

\subsection{Concentration Mechanisms }

Our arguments require similarity among the competitors. Concentrated populations are produced naturally by a variety of selection dynamics. Strict NE that maximize average performance on their local neighborhood may act as attractors on the interior of the trait space. Selection pressure towards a boundary of the trait space may also induce concentration when competition promotes extremal values of some traits, e.g. speed among runners. Then, the distribution of strongest competitors will concentrate at that physiological barrier. The game theory literature adopts a standard story. First, a static equilibrium concept is introduced, then it is shown that, under a chosen selection dynamic, population distributions initialized near enough to the equilibrium will converge, under an appropriate topology, to a monomorphism (delta distribution) at the equilibrium. Thus, widely proved statements regarding the stability of monomorphic equilibria establish concentration mechanisms (see \cite{cabrales2000stochastic,Cressman_b,Cressman_c,foster1990stochastic,fudenberg1992evolutionary,imhof2005long,kandori1993learning,oechssler2001evolutionary}).

For example, the replicator equation is commonly used to model strategy evolution in evolutionary game theory (see all but \cite{kandori1993learning} of the previous list). It models systems where competitive payout translates to per capita population growth rate. For any strategy $x$, the rate of change in the proportion of the population playing strategy $x$ depends on the difference between the expected payout to $x$ and the average payout of all available strategies. The portion of the population outperforming the average grows while the rest shrinks. This dynamic arises naturally in biological contexts, and in economics as the limit of a reinforcement learning process \cite{borgers1997learning}.\footnote{For a review of replicator equation dynamics, see \cite{Cressman_a}. Extensions to replicator-like dynamics are reviewed in \cite{fudenberg1998theory}. In particular, monotone and myopic adjustment dynamics, which relax the replicator dynamic while requiring that, in some sense, strategies with high expected payouts grow faster than strategies with low payouts, eliminate strictly dominated strategies. These processes promote concentration in support to a subset of strategies that survive a process of iterated dominance \cite{samuelson1992evolutionary,hofbauer1996evolutionary}.  } The Folk Theorem of Evolutionary Game Theory implies that all strict Nash equilibria of a matrix game with finitely many pure strategies are asymptotically stable under the replicator dynamic. Furthermore, all populations in the  interior of the strategy space evolve to a Nash equilibrium (NE) \cite{Hofbauer2003folk}. This result extends to evolutionary stable strategies (ESS); any strategy that is both a NE and is uninvasible  \cite{Maynard_Smith}. An ESS on the interior of the strategy space is a globally asymptotically stable rest point for the replicator equation, while an ESS on the boundary is locally asymptotically stable. Convergence to an ESS or NE implies concentration in the space of mixed strategies.

Yet stronger equilibrium notions, such as continuously stable strategies (CSS) \cite{Eshel}, or neighborhood invader strategies (NIS) \cite{Apaloo}, follow. Stronger equilibrium notions are introduced to guarantee the stability of monomorphisms (delta distributions) in continuous trait spaces under adaptive dynamics \cite{dieckmann1996dynamical,leimar2005multidimensional}, or continuous space replicator equations \cite{cressman2004dynamic,Cressman_b,Cressman_c,oechssler2001evolutionary}.\footnote{Note that, even if the set of pure strategies of a game is finite, the space of possible mixed strategies is continuous, so the dynamics of populations adopting a mixed strategy is itself a continuous trait space. Concentration in this space corresponds to a population of individuals with similar, albeit stochastic, behavior rules. These rules determine expected payout, so concentration in the mixed strategy space to a single mixed strategy implies our main convergence result.} Convergence to a monomorphism implies concentration in the distribution of traits of individuals in a population. 

Similar results even extend to evolutionary models subject to small random perturbations in the payouts, low levels of mutation, or background immigration \cite{fudenberg1998theory}.\footnote{As an extreme example, if mutation rates are very small relative to selection dynamics, then mutation is rare, and most mutations either fixate or die out before another enters. In this case, ecological processes (population dynamics) occur on a  much faster time scale than evolutionary dynamics \cite{cressman2004coevolution}, so, at any time, populations will remain close to monomorphic in any given trait. The motion of the monomorphism is governed by a random walk, which may be modeled with a Markov process \cite{sella2005application}, or via adaptive dynamics \cite{abrams2001modelling,dieckmann1996dynamical,leimar2005multidimensional}. Similar models may arise in learning processes which iteratively sample small variations on a chosen strategy, compare those variations, then select one of the variations.} In the presence of noise, it is usually shown that an ergodic stationary distribution converges, in the limit of small noise, to a delta distribution \cite{kandori1993learning,foster1990stochastic}. Rates of convergence, iterated elimination of dominated strategies, and equilibrium selection can be established rigorously in this context using approaches adapted from Friedlin and Wentzell \cite{fudenberg1998theory,foster1990stochastic}, or Kandori and Rob \cite{kandori1993learning}. For examples see \cite{cabrales2000stochastic,foster1990stochastic,fudenberg1992evolutionary,imhof2005long}. In those cases, low but finite levels of noise lead to highly concentrated distributions.\footnote{In these cases, the stochastically stable set (limit of the support of the concentrating distribution), may depend on apparently fine details regarding the noise implementation, and may or may not correspond to static equilibrium concepts such as an ESS.} Similarly, low levels of mutation or immigration may sustain steady states near to a monomorphism \cite{cabrales2000stochastic,fudenberg1992evolutionary}. 

We save mechanism specific analysis and conditions for concentration for future work. The former question is model specific. The latter is well studied.

%\cc{Rephrase paragraph to account for more rigorous genetic drift examples: } Classically, the replicator equation does not account for genetic drift caused by mutation. Genetic drift can be introduced by introducing a small mutations upon replication. When the drift is small, concentrated distributions may form near evolutionarily stable strategies, which are attracting in the deterministic dynamics. The center of that distribution will be near the Nash equilibrium, which outcompetes opponents that deviate too far from it. Therefore, if the ESS is located in the interior of the strategy space, we would expect the competitors to concentrate in a small subregion of all available strategies. For more details, see \cite{fudenberg1998theory}. 

\subsection{Overview}

Why should similarity promote transitivity?  If performance is smooth, then, on small enough regions in trait space, it is approximately linear. 

All linear functions of $x$ and $y$ satisfying $f(x,y) = - f(y,x)$ admit some rating function $r(x)$ such that $f(x,y) = r(x) - r(y)$. Thus, all linear performance functions are perfectly transitive, so, up to a linear approximation, performance appears transitive on small neighborhoods. Convergence to transitivity is controlled by higher order terms, i.e.~quadratic expansions. Interactions between distinct attributes of distinct agents produce cycles at second order. Second order terms vanish faster than first order terms, so the cyclic component vanishes faster than the transitive component.

These observations ground all of our results. They also explain systematic differences in the convergence rates to transitivity depending on where the distribution concentrates, the dimension of the trait space, and the structure of the performance function. First, if $f$ is rough, then it is highly nonlinear, so it only appears linear on small neighborhoods in trait space. Thus, the degree of similarity required depends on the smoothness of $f$. The smoother $f$, the more transitive it appears. Second, suppose the trait distribution concentrates about a location where the local linear approximation to $f$ is nearly flat, and thus near zero. Such a situation may occur at the end of an evolutionary or training process where the population approaches a local maximum in some effective rating function (c.f.~\cite{Cressman_c}). Then convergence may require very small neighborhoods, or may not occur at all. Third, if competition only depends on a single traits, or traits do not interact (at second order), then cycles can only enter via third or higher order terms, so convergence to transitivity occurs unusually quickly, and on larger neighborhoods in trait space.

The rest of the paper expands these ideas in detail. Section \ref{sec: results} develops the underlying theory rigorously. We present out main concentration theorem in section \ref{sec: concentration generic}. The theorem depends on the trait-performance theorem introduced in \cite{strang2022network} and restated in section \ref{sec: trait performance}. The trait performance theorem establishes that the expected size of $f_c$ is controlled by the correlation, $\rho$, in the performance of an agent against two randomly chosen opponents. The larger the correlation, the more transitive competition.  If $\rho = 1/2$, then competition is perfectly transitive. Our concentration theorem establishes that, under appropriate smoothness and concentration assumptions, $\rho$ differs from $1/2$ by a small quantity $\epsilon$ which converges to zero as the trait distribution concentrates. Thus similarity suppresses cyclicity.  A reader who is not interested in the analysis may safely skip the intermediate sections. For concision, most proofs and supplementary calculations are provided in the SI (see Appendix \ref{app: supplemental proofs}). 

Section \ref{sec: numerics} documents a pair of numerical experiments which support our theory. We develop a phenomenological evolution model based on simple genetic training algorithms, and apply it to a series of games. These include a set of illustrative bimatrix games, and a sequence of randomly generated games with tuneable structure designed to demonstrate generality. In almost all cases we observe convergence to transitivity via trait concentration, with convergence rates that match our theoretical predictions.

%%%%%%%%%%%%%%%%%%%%%%%%%%%%%%%%%%%%%%%%%%%%%%%%%%%%%%%%%%%%%%%%%%
\section{Results} \label{sec: results}

\subsection{Local Expansion of Performance} \label{sec: taylor expansion}

To study convergence to transitivity, we need a local model for competition. We focus on local quadratic models, which provide the simplest, sufficiently general, framework. Quadratic performance functions also arise naturally (c.f.~\cite{imhof2005long,oechssler2001evolutionary}). For example, the expected payout of mixed strategy $p$ against mixed strategy $q$ in any zero-sum bimatrix game is a quadratic performance function $f(p,q)$. Quadratic models also provide a general theory for smooth performance functions on small neighborhoods via Taylor approximation \cite{cressman2004coevolution, Cressman_b, Cressman_c}.

Suppose that $f(x,y)$ is continuously second differentiable at all $x = y = z \in \Omega$. In addition suppose that the Taylor expansion of $f(x,y)$ about $x = y = z$ converges on a ball with finite radius for all $z \in \Omega$, or, for any $z$, there exists a ball of finite radius containing $z$ where the errors in the second order Taylor expansion of $f$ about $z$ can be bounded above by a power series whose lowest order terms are cubic. In either case, $f(x,y)$ can be approximated on local neighborhoods of $z$ by its second order Taylor expansion. Let $\nabla_x$ denote the gradient with respect to the traits of the first competitor and $\nabla_y$ denote the gradient with respect to the traits of the second competitor. Let $H(x,y)$ denote the Hessian of the performance function. Then $H$ can be written in the block form:
\begin{equation}
    H(x,y) = \left[ \begin{array}{cc} H_{xx}(x,y) & H_{xy}(x,y) \\ H_{yx}(x,y) & H_{yy}(x,y) \end{array} \right]
\end{equation}
where the subscripts denote which partials are contained in the block. If there are $T$ traits then each block is $T \times T$ and $H_{xx}$ contains all second order partials in the traits of the first competitor, $H_{yy}$ contains all second order partials in the traits of the second competitor, and $H_{xy} = H_{yx}^{\intercal}$ store the cross partials.

Let $z$ be some trait vector near $x$ and $y$. Then, to second order:
\begin{equation}
\begin{aligned} \label{eqn: quadratic not simplified}
    f(x,y) \simeq & f(z,z) + \nabla_x f(z,z)^{\intercal} (x - z) + \nabla_y f(z,z)^{\intercal} (y - z) + \hdots \\ &  + \frac{1}{2} \left[  (x - z)^{\intercal},  (y - z)^{\intercal} \right] \left[ \begin{array}{cc} H_{xx}(z,z) & H_{xy}(z,z) \\ H_{yx}(z,z) & H_{yy}(z,z) \end{array} \right] \left[ \begin{array}{c} x - z \\ y - z \end{array} \right] + \mathcal{O}((x -z, y -z)^3).
\end{aligned}
\end{equation}

The Taylor expansion simplifies since $f$ is alternating, $f(x,y) = - f(y,x)$. Therefore $f(z,z) = - f(z,z)$ so $f(z,z) = 0$ for all $z$. Stronger alternation requirements follow. Let $\alpha, \beta$ be multi-indices, and let $
    \partial_x^{\alpha} = \partial_{x_1}^{\alpha_1} \partial_{x_2}^{\alpha_2} \hdots \partial_{x_T}^{\alpha_T}$.
    %\partial_y^{\beta} = \partial_{y_1}^{\beta_1} \partial_{y_2}^{\beta_2} \hdots \partial_{y_T}^{\beta_T}$
Then:
\begin{equation} \label{eqn: fairness derivative requirement}
    \partial_{x}^{\alpha} \partial_{y}^{\beta} f(x,y)|_{x=u,y=w} =  - \partial_{x}^{\beta} \partial_y^{\alpha} f(x,y)|_{x=w,y=u}.
\end{equation} 

Applying Equation \ref{eqn: fairness derivative requirement} at $u = w = z$ yields:
\begin{equation} \label{eqn: alternating derivatives}
    \partial_{x}^{\alpha} \partial_{y}^{\beta} f(z,z) =  - \partial_{x}^{\beta} \partial_y^{\alpha} f(z,z).
\end{equation}
where $\partial_x$ and $\partial_y$ denote partials with respect to the first and second competitors respectively. 

Equation \ref{eqn: alternating derivatives} has interesting implications for the second order Taylor expansion of $f$. In particular:
\begin{equation} \label{eqn: alternating gradient and Hessian}
    \begin{aligned}
        & \partial_{x_i} f(z,z) = - \partial_{y_i} f(z,z) & \implies & \nabla_{x} f(z,z) = - \nabla_y f(z,z) \\
        & \partial_{x_i} \partial_{x_j} f(z,z) = - \partial_{y_i} \partial_{y_j} f(z,z) & \implies & H_{xx}(z,z) = - H_{yy}(z,z) \\
        & \partial_{x_i} \partial_{y_j} f(z,z) = - \partial_{y_i} \partial_{x_j} f(z,z) & \implies & H_{xy}(z,z) = -H_{yx}(z,z)  \\
    \end{aligned}
\end{equation}
which follow from letting $(\alpha = e_i, \beta = 0)$, $(\alpha = e_i + e_j, \beta = 0)$, and $(\alpha = e_i, \beta = e_j)$ where $e_k$ denotes the $T \times 1$ indicator vector for trait $k$.

Since $H(z,z)$ is a Hessian, it must be symmetric. Thus $H_{xx} = H_{xx}^{\intercal} = - H_{yy}^{\intercal} = - H_{yy}$ and $H_{xy} = H_{yx}^{\intercal} = - H_{xy}^{\intercal} = - H_{yx}$. Then the diagonal blocks $H_{xx}, H_{yy}$ are both symmetric while the off-diagonal blocks $H_{xy}, H_{yx}$ are skew symmetric. That is,  $H_{xy} = - H_{xy}^{\intercal}$, and  $ H_{yx} = - H_{yx}^{\intercal}.$

The alternating structure of the derivatives simplifies the quadratic approximation to $f$. To second order, performance is a difference in a local rating functions $r(\cdot|z)$, plus a term coupling distinct traits. Specifically:
\begin{equation} \label{eqn: quadratic approximation to performance}
    f(x,y) \simeq r(x|z) - r(y|z) + (x - z)^{\intercal} H_{xy}(z,z) (y - z) + \mathcal{O}((x - z, y - z)^3).
\end{equation}
where:
\begin{equation} \label{eqn: local rating function}
    r(x|z) = \nabla_x f(z,z)^{\intercal} (x - z) + \frac{1}{2}(x - z)^{\intercal} H_{xx}(z,z) (x - z). 
\end{equation}
See Appendix \ref{app: supplemental proofs} for the details. Note that this performance function is similar to the Taylor expansion of the fitness function in \cite{Cressman_c}, except our function is skew symmetric, not symmetric.

There are two morals here. First, at first order, $f$ is the difference in a pair of affine local rating functions:
\begin{equation} \label{eqn: first order rating function}
    f(x,y) \simeq r^{(1)}(x|z) - r^{(1)}(y|z) + \mathcal{O}((x - z,y-z)^{2}) \text{ where } r^{(1)}(x,z) = \nabla_x f(z,z)^{\intercal} (x - z).
\end{equation}

When performance equals a rating difference, any associated competitive network is perfectly transitive. Thus, on small enough neighborhoods, competition will be close to perfectly transitive. Any remaining intransitivity must enter at second order, so, as the neighborhood concentrates, competition should approach perfect transitivity.

Second, the block decomposition of $H$ breaks into transitive and cyclic parts. The on-diagonal blocks, $H_{xx}$ and $H_{yy}$, introduce curvature to the local rating functions, so are perfectly transitive. Thus, the off-diagonal blocks $H_{xy}$ are, locally, the only source of cyclicity.

The skew-symmetry of $H_{xy}$ implies that all of its diagonal entries are zero. Therefore:

\vspace{2mm}
\textbf{Lemma 1 (Perfect Transitivity to Second Order): } \textit{If the trait space is one-dimensional ($T = 1$), or the off-diagonal blocks of the Hessian $H(z,z)$ are diagonal, then the local quadratic approximation to performance is perfectly transitive. }
\vspace{2mm}

\textbf{Proof} The off-diagonal block of the Hessian is skew-symmetric, so all of its diagonal entries are zero. It follows that if the trait-space is one dimensional, or the block is diagonal, then $H_{xy} = 0$. Then, by equation \ref{eqn: quadratic approximation to performance}, the local quadratic approximation to performance is a difference in a pair of rating functions. $\blacksquare$

To see why the diagonal requirement on $H_{xy}$ is interesting, consider a performance function of the form:
\begin{equation} \label{eqn: noninteracting traits}
    f(x,y) = \sum_{j=1}^{T} g_j(x_j,y_j)
\end{equation}
where $g_j$ are a set of single-trait performance functions (alternating in $x_j,y_j$). %For example, if performance represents log-odds of victory then equation \ref{eqn: noninteracting traits} requires that the odds a competitor with traits $x$ beats a competitor with traits $y$ can be written as a product of functions per individual trait. 
In this case there is no interaction between different traits. Instead, performance consists of a series of trait-by-trait comparisons. Performance functions of this kind are convenient for numerical tests since they are easy to construct, however, they are unusually transitive (see Lemma 1). 

 Therefore, to second order, intransitivity on small neighborhoods requires non-zero off-diagonal terms in $H_{xy}$ that couple distinct traits of the competitors. Thus, intransitivity on small neighborhoods arises from comparisons of distinct traits. When these interactions vanish, intransitivity only enters at third order, producing unusually fast convergence to transitivity.

\subsection{Trait-Performance Theory}

How transitive/cyclic are competitive networks whose competitors are sampled from small local neighborhoods? We answer that question using the trait-performance theorem established in \cite{strang2022network}.

\subsubsection{Trait Performance Theorem} \label{sec: trait performance}

In a trait-performance model, competitive events are mediated by the competitor traits, which are sampled i.i.d from a trait distribution that models the demographics of the population. Let $X,Y,W \in \mathbb{R}^T$ denote the traits of three different competitors, where $T$ is the number of relevant traits, and let $\pi_x$ denote the trait distribution they are sampled from. Assume that the advantage one competitor possesses over another is independent of their location in the network and can be expressed as a deterministic function of their traits. Then there must exist a performance function $f(x,y)$, satisfying $f(x,y) = - f(y,x)$, that returns the advantage a competitor with trait $x$ possesses over competitor with trait $y$. 

\vspace{2mm}

\label{thm: expected transitivity and correlation coefficient}
\textbf{Theorem 1: (Trait Performance)} \textit{Let $\mathcal{G}$ be a competitive network with $V$ vertices and $E$ edges, where the traits of each competitor are drawn independently from $\pi_x$, and the edge flow is defined by $F_k = f(X(i(k)),X(j(k)))$ where $f(x,y)$ is an alternating function. Then the covariance $\mathbb{V}[F]$ of the edge flow has the form:}
\begin{equation} \label{eqn: generic covariance}
\mathbb{V}[F] = \sigma^2 \left[I + \rho \left(G G^T - 2 I \right) \right]
\end{equation}
\textit{where $\sigma^2$ is the variance in $F_k$ for arbitrary $k$, $\rho$ is the correlation coefficient between $f(X,Y)$ and $f(X,W)$ for $X,Y,W$  drawn i.i.d from $\pi_x$, and $G$ is the edge-incidence matrix for $\mathcal{G}$.}
		
\textit{Moreover:}
\begin{equation} \label{eqn: trait performance}
\mathbb{E} \left[\frac{1}{E}||F||^2 \right] = \sigma^2  \xrightarrow{\text{decompose}} \left\{\begin{aligned}
	& \mathbb{E} \left[\frac{1}{E}||F_t||^2 \right] = \sigma^2 \left[\frac{(V-1)}{E} + 2 \rho \frac{L}{E} \right] \\
	& \mathbb{E} \left[\frac{1}{E}||F_c||^2 \right] = \sigma^2 \left(1 - 2 \rho \right) \frac{L}{E} \end{aligned} \right.
\end{equation}
\textit{where $L = E - (V - 1)$ is the dimension of the cycle space of $\mathcal{G}$. }

\textit{The size of the transitive component is monotonically increasing in $\rho$, and the size of the cyclic component is monotonically decreasing in $\rho$, where $\rho$ ranges from $0$ to $1/2$. If $\rho = 1/2$, then competition is perfectly transitive.}

\vspace{2mm}

Theorem 1 states that, when performance is a function of randomly sampled traits, the expected degree of intransitivity depends only on the network dimensions and a pair of local performance statistics, $\sigma^2$ and $\rho$ (see Equation \ref{eqn: trait performance}). The actual network topology does not influence the expectation. Thus we do not need to consider specific random graphs, only $\sigma^2$ and $\rho$.

The correlation coefficient $\rho$ controls the expected relative sizes of transitive and cyclic competition. The correlation coefficient can be expanded:
\begin{equation} \label{eqn: rho integral}
    \rho = \frac{\mathbb{V}_{X}[\mathbb{E}_{Y}[f(X,Y)]]}{\mathbb{V}_{X,Y}[f(X,Y)]} = \frac{\int_{\Omega} \left(\int_{\Omega} f(x,y) \pi(y) dy \right)^2 \pi_x(x) dx}{\int_{\Omega} \int_{\Omega} f(x,y)^2 \pi_x(y) \pi_x(x) dy dx}
\end{equation}
since $X$ and $Y$ are drawn i.i.d. and $f$ is alternating, thus $\mathbb{E}[f(X,Y)] = 0$ (see \cite{strang2022network}).
The correlation $\rho$ is the variance in the expected performance, $\mathbb{E}_{Y}[f(X,Y)]$ conditioned on the traits $X$, normalized by the variance in performance. Thus, variance in expected performance promotes transitive competition. Variance in expected performance promotes transitive competition since it implies that we frequently sample some competitors who perform well against most opponents, and some who perform poorly against most opponents.

We use $\rho$ to study the how trait concentration promotes transitivity and suppresses cyclicity. In the next section we will introduce a small quantity $\epsilon$ that controls how far $\rho$ is from $1/2$, and thus the expected sizes of the transitive and cyclic components. We show that $\epsilon$ usually vanishes as trait distributions concentrate, so $\epsilon$ links the breadth of the trait distribution to the expected structure of competition.

\subsubsection{Bounding the Correlation Coefficient} \label{sec: correlation bounds}

The correlation coefficient $\rho$ controls the expected relative sizes of the components. What is $\rho$ given a quadratic, or nearly quadratic, performance function?

To answer these questions, write:
\begin{equation}
    f(x,y) = r(x|z) - r(y|z) + h(x,y|z)
\end{equation}
where $r(x,|z)$ are local rating functions based on the expansion about $z$. We can always shift $r(x|z)$ by a constant, since $r(x|z) + c - (r(y|z) + c) = r(x|z) - r(y|z)$. Thus we are free to center $r(x|z)$ so that $\mathbb{E}[r(X|z)] = 0$. The remaining terms, $h(x,y|z)$ accounts for the higher order or intransitive behavior that is not captured by the local rating function.  

The correlation coefficient $\rho$ is given by a ratio (see Equation \ref{eqn: rho integral}). The numerator is the uncertainty in the expected performance of a competitor with traits $x$. The denominator is the uncertainty in performance between two randomly chosen agents. To simplify notation, we suppress the $z$ dependence for now.

Then (see Appendix \ref{app: supplemental proofs}) the correlation coefficient is given by:
\begin{equation} \label{eqn: rho in expectations}
    \rho = \frac{\mathbb{V}_{X}[r(X)] + 2 \mathbb{E}_{X,Y}[r(X) h(X,Y)] + \mathbb{E}_{X}\left[ \mathbb{E}_{Y}[h(X,Y)]^2 \right] }{2 \mathbb{V}_{X}[r(X)] + 4 \mathbb{E}_{X,Y}[r(X) h(X,Y)] + \mathbb{E}_{X,Y}[h(X,Y)^2]}
\end{equation}
The first two terms in the denominator of Equation \ref{eqn: rho in expectations} are exactly twice the corresponding terms in the numerator. The ratio of the last two terms is of the same form as the ratio that defined $\rho$ (see \ref{eqn: rho integral}). It follows immediately that $\rho \in [0,1/2]$, and, if performance is perfectly transitive, $h(x,y) = 0$, so $\rho = 1/2$. 

To bound $\rho$ from below note that $\mathbb{E}_{X}\left[ \mathbb{E}_{Y}[h(X,Y)]^2 \right] \geq 0$ so:
$$
    \rho \geq \frac{\mathbb{V}_{X}[r(X)] + 2 \mathbb{E}_{X,Y}[r(X) h(X,Y)]}{2 \mathbb{V}_{X}[r(X)] + 4 \mathbb{E}_{X,Y}[r(X) h(X,Y)] + \mathbb{E}_{X,Y}[h(X,Y)^2]}
$$

Define $\epsilon$: 
\begin{equation} \label{eqn: epsilon}
    \epsilon = \frac{\mathbb{E}_{X,Y}[h(X,Y|z)^2]}{2 \mathbb{V}_{X}[r(X|z)] + 4 \mathbb{E}_{X,Y}[r(X|z) h(X,Y|z)]}
\end{equation}

Then:
\begin{equation} \label{eqn: epsilon bound on rho}
    \rho \geq \frac{1}{2} \frac{1}{1 + \epsilon}.
\end{equation}

Note that, $
1/2 - \rho \leq \frac{1}{2}\left(1 - \frac{1}{1 + \epsilon} \right) = \frac{1}{2} \frac{\epsilon}{1 + \epsilon} \leq \frac{1}{2} \epsilon
$. Then $1 - 2 \rho \leq \epsilon$ so $\mathbb{E}[||F_c||^2]$ is $\mathcal{O}(\epsilon)$ while $\mathbb{E}[||F_t||^2]$ is $\mathcal{O}(1)$. So, the smaller $\epsilon$ the more transitive and less cyclic competition. Thus $\epsilon$ acts as a relevant small quantity that bounds the size of the cyclic component of competition. We will show in section \ref{sec: trait concentration} that $\epsilon$ typically vanishes as a trait distribution concentrates, so it controls rate of convergence to transitivity.

\begin{figure}[t]
    \centering
    \includegraphics[scale=0.3]{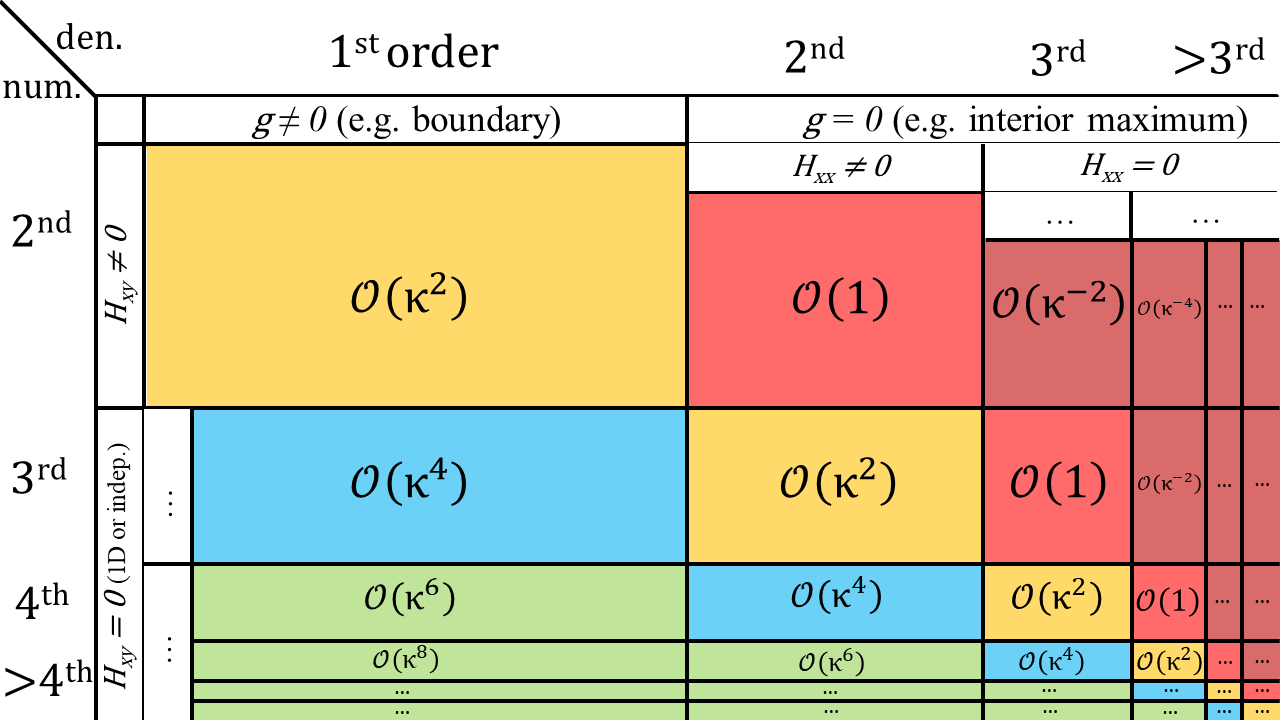}
    \caption{Predicted convergence rates to transitivity under varying conditions on the partial derivatives of performance at the trait distribution centroid. The denominator (column) and numerator (row) reference Equation \ref{eqn: epsilon} and the colors correspond to different rates of convergence. In general, $H_{xy} \neq 0$ unless the trait space is one-dimensional, or distinct traits do not interact (see Lemma 1). Similarly, the gradient $g = \nabla_x f$ is generically nonzero, though evolutionary processes may converge towards local maxima in an effective rating function on the interior where the gradient vanishes. Processes concentrating at the boundary will generically have nonzero gradient applying selection pressure into the boundary. } %In our experimental setup, we observe $H_{xy} = 0$ when we have one-dimensional trait distributions like bimatrix games, or with multi-dimensional trait distributions with no cross-interactions among pairs. We observe zero gradient in the case of prisoner's dilemma, along with some trials of stag hunt, leading to a predicted convergence rate of $\mathcal{O}(\kappa^4)$ in those cases. With other stag hunt trials, we observe nonzero gradients leading to a convergence rate of $\mathcal{O}(\kappa^2)$. In chicken, both the first-order numerator and second-order denominator could go to zero at different points, which could lead to dominance by the higher-order terms.}
    \label{fig:Convergence rate flowchart}
\end{figure}

Figure \ref{fig:Convergence rate flowchart} illustrates the theoretical predicted rates of convergence towards transitivity based on which order derivatives of $f$ are nonzero. As observed in Theorem 2, the order at which derivatives vanish controls the rate of convergence with respect to central moments of the trait distribution. Taylor expanding Equation \ref{eqn: epsilon} produces a rational function whose numerator and denominator are polynomials including only even order moments. The table predicts rates of convergence to transitivity based on the lowest order term in the numerator and denominator, borrowing the bounds on $\rho$ from Equation \ref{eqn: epsilon bound on rho}. 

In the special case when performance is quadratic $\epsilon$ simplifies and predicts $\rho$ exactly:

\vspace{2mm}

\textbf{Lemma 2: (Quadratic Performance and $\epsilon$)} 
\textit{If the performance function is quadratic, then:} 
\begin{equation} \label{eqn: quadratic epsilon}
\rho = \frac{1}{2} \frac{1}{1 + \epsilon}, \textit{ and } 
    \epsilon = \frac{\mathbb{E}_{X,Y}[h(X,Y|z)^2]}{2 \mathbb{V}_{X}[r(X|z)]}.
\end{equation}
\textit{so:}
\begin{equation} \label{eqn: trait performance by epsilon}
\mathbb{E} \left[\frac{1}{E}||F||^2 \right] = \sigma^2  \xrightarrow{\text{decompose}} \left\{\begin{aligned}
	& \mathbb{E} \left[\frac{1}{E}||F_t||^2 \right] = \sigma^2 \left[\frac{(V-1)}{E} + \frac{1}{1 + \epsilon} \frac{L}{E} \right] = \mathcal{O}(1) \\
	& \mathbb{E} \left[\frac{1}{E}||F_c||^2 \right] = \sigma^2 \frac{\epsilon}{1 + \epsilon} \frac{L}{E} = \mathcal{O}(\epsilon) \end{aligned} \right.
\end{equation}

\vspace{2mm}

The proof is provided in Appendix \ref{app: supplemental proofs}. It follows by demonstrating that $\mathbb{E}_{X}\left[ \mathbb{E}_{Y}[h(X,Y)]^2 \right] = 0$ and $\mathbb{E}_{X,Y}[r(X|z) h(X,Y|z)] = 0$ when performance is quadratic. We use this simplification to study convergence to transitivity by adopting quadratic approximations in a concentration limit.

The next section expands $\epsilon$ and $\rho$ in terms of the moments of the trait distribution, provided performance is quadratic. We return to quadratic approximation in the concentration limit in Section \ref{sec: trait concentration}.

\subsubsection{Computing Epsilon} \label{sec: epsilon calculation}

Suppose $f$ is quadratic. Then $h(x,y|z) = (x - z)^{\intercal} H_{xy}(z,z) (y - z)$ where $z$ is chosen so that $z = \mathbb{E}_X[X]$. Since performance is quadratic, the Hessian is independent of $z$, so we suppress the $z$ dependence in $H$. Then $\epsilon$ is determined by Equation \ref{eqn: quadratic epsilon} where:
$$
\begin{aligned}
& r(x) = \nabla_x f(z,z)^{\intercal} (x - z) + \frac{1}{2} (x - z)^{\intercal} H_{xx} (x - z) - \bar{r} \\
& h(x,y) = (x - z)^{\intercal} H_{xy} (y - z). 
\end{aligned}
$$
and where the constant $\bar{r}$ is chosen so that $\mathbb{E}_X[r(X)] = 0$.\footnote{Specifically:
\begin{equation}
\begin{aligned}
    \bar{r} = & \mathbb{E}_{X}[r(X)] = \frac{1}{2} \sum_{i,j} {H_{xx}}_{i,j} \text{Cov}[X]_{i,j}  = \text{trace}[H_{xx} \text{Cov}[X]] = \langle H_{xx}, \text{Cov}[X] \rangle
\end{aligned}
\end{equation}
where $\langle A, B \rangle$ is the matrix inner product $\sum_{i,j} a_{i,j} b_{i,j}$.}

To compute $\epsilon$ explicitly, expand it in terms of the central moments of $\pi_x$. Let $\text{Cov}[X]$ stand for the covariance, $\text{Th}[X]$ stand for the tensor of third order central moments, and $\text{Fo}[X]$ the tensor of fourth order central moments. To express tensor products we adopt Einstein summation notation.\footnote{Greek superscripts and subscripts denote dimensions of a tensor. Perform elementwise products of all matching Greek super/subscripts, and sum across dimensions where the same Greek letter appears as both scripts. For example, a standard matrix vector product would be written $A^{\alpha}_{\beta} x_{\alpha}$.}

Then, the variance in the quadratic local rating function is (see Appendix \ref{app: supplemental proofs}):
\begin{equation}\label{eqn: variance in ratings}
\begin{aligned}
    \mathbb{V}_X[r(X)]  = & \nabla_x f(z,z)^{\alpha} \nabla_x f(z,z)^{\beta} \text{Cov}[X]_{\alpha \beta} + \nabla f(z,z)^{\alpha} H_{xx}^{\mu \nu} \text{Th}[X]_{\alpha \mu \nu} + \hdots \\ & \hdots + \frac{1}{4} H_{xx}^{\alpha \beta} H_{xx}^{\mu \nu} \left( \text{Fo}[X]_{\alpha \beta \mu \nu} - \text{Cov}[X]_{\alpha \beta} \text{Cov}[X]_{\mu \nu} \right) 
\end{aligned}
\end{equation}

The numerator of Equation \ref{eqn: rho integral} is (see Appendix \ref{app: supplemental proofs}):
\begin{equation} \label{eqn: expected h squared}
    \mathbb{E}_{X,Y}[h(X,Y)^2] = H_{xy}^{\alpha \beta} H_{xy}^{\mu \nu} \text{Cov}[X]_{\alpha \mu} \text{Cov}[X]_{\beta \nu}
\end{equation}

Then:
\begin{equation} \label{eqn: epsilon explicitly}
\epsilon = \frac{1}{2}\ltfrac{H_{xy}^{\alpha \beta} H_{xy}^{\mu \nu} \text{Cov}[X]_{\alpha \mu} \text{Cov}[X]_{\beta \nu}}{ \nabla_x f(z,z)^{\alpha} \nabla_x f(z,z)^{\beta} \text{Cov}[X]_{\alpha \beta} + \nabla f(z,z)^{\alpha} H_{xx}^{\mu \nu} \text{Th}[X]_{\alpha \mu \nu} + \frac{1}{4} H_{xx}^{\alpha \beta} H_{xx}^{\mu \nu} \left(\text{Fo}[X]_{\alpha \beta \mu \nu} - \text{Cov}[X]_{\alpha \beta} \text{Cov}[X]_{\mu \nu}  \right)}. 
\end{equation}

In special cases, we can simplify Equation \ref{eqn: epsilon explicitly}. We use these cases to predict $\epsilon$ in our numerical tests.

Suppose that the trait distribution is not skewed. Then the third order central moments vanish and:
\begin{equation}
    \epsilon = \frac{1}{2}\frac{H_{xy}^{\alpha \beta} H_{xy}^{\mu \nu} \text{Cov}[X]_{\alpha \mu} \text{Cov}[X]_{\beta \nu}}{\nabla_x f(z,z)^{\alpha} \nabla_x f(z,z)^{\beta} \text{Cov}[X]_{\alpha \beta} + \frac{1}{4} H_{xx}^{\alpha \beta} H_{xx}^{\mu \nu} \left(\text{Fo}[X]_{\alpha \beta \mu \nu} - \text{Cov}[X]_{\alpha \beta} \text{Cov}[X]_{\mu \nu} \right) }. 
\end{equation}

Suppose, in addition, the trait distribution is multivariate normal as may arise due to natural phenotypic variation satisfying a central limit theorem, or as the result of a selection process. Examples are used in  \cite{cressman2004coevolution, Cressman_c}, where it is shown that normally distributed populations remain normally distributed under a continuous trait replicator dynamic applied to a quadratic performance function. When normal, the fourth order central moments are determined by the covariance in the trait distribution. Given $X \sim \mathcal{N}(z,\Sigma)$:
\begin{equation} \label{eqn: epsilon normal}
\epsilon = \frac{H_{xy}^{\alpha \beta} H_{xy}^{\mu \nu} \Sigma_{\alpha \mu} \Sigma_{\beta \nu}}{2 \nabla_x f(z,z)^{\alpha} \nabla_x f(z,z)^{\beta} \Sigma_{\alpha \beta} +  H_{xx}^{\alpha \beta} H_{xx}^{\mu \nu} \Sigma_{\alpha \mu} \Sigma_{\beta \nu} }.
\end{equation}

Suppose that the trait distribution reflects the long time limit of some evolutionary process, and the trait distribution is reasonably concentrated and unimodal. Then, when the distribution is concentrated at some $z$, the average of the gradient in the quadratic local rating function is $\nabla_x f(z,z)$. As long as this gradient is nonzero, the distribution should not remain stationary, since $\nabla_x f(z,z)$ identifies a direction that would improve a competitor's expected performance. Thus, we might expect an evolutionary process to move a concentrated trait distribution until its centroid $z$ achieves $\nabla_x f(z,z) = 0$ \cite{Cressman_b,marrow1996evolutionary}. Alternatively, if the region of admissable traits is bounded, then the distribution could reach a steady state where the gradient points into the boundary, thus trapping the distribution (c.f.~the war of attrition example in \cite{oechssler2001evolutionary}). 

These arguments can be formalized by the canonical equation of adaptive dynamics. Adaptive dynamics studies the motion of highly concentrated populations, where concentration is justified by rare, small mutations. Then populations remain near to monomorphic (delta distributed), and are directed by a vector field, $\frac{d}{dt} \bar{x}(t) = C(\bar{x}(t),t) g(\bar{x}(t))$ where $\bar{x}(t)$ is the population centroid, $g(z) = \nabla_x f(x,y)|_{x=y=z}$ is the local gradient in performance evaluated in the neighborhood of $x$, and $C(x,t)$ is a symmetric p.s.d~matrix representing either the population covariance, or variation introduced by mutation \cite{abrams2001modelling,Cressman_c,dieckmann1996dynamical,leimar2005multidimensional,meszena2002evolutionary,zeeman1981dynamics}. The gradient $g(z)$ defines a locally linear model for performance, and points in the direction of fastest improvement against individuals drawn from a small neighborhood of $z$.  The same dynamics can be justified by studying the motion of the centroid of a distributed population (see \cite{cressman2004coevolution}). Then, provided $C(x_*,t)$ is not singular, $x_*$ is not an equilibrium for $\bar{x}(t)$ unless the gradient vanishes at $x_*$, or $x_*$ lies on the boundary of the trait space at a point where the gradient points into the boundary \cite{Cressman_c,Cressman_b}. In the former case the Hessian $H_xx$ must be negative definite at $x_*$ to ensure stability \cite{Cressman_c}. See \cite{Cressman_c,leimar2005multidimensional} for further details on the relation between negative definiteness and equilibrium concepts in game theory. In the latter case, the gradient must also have no projection onto any vector in the tangent space to the boundary at $x_*$, otherwise $\bar{x}(t)$ will drift in that direction.  

These two cases lead to different convergence behavior towards transitivity. Recall that transitive competition dominates on small neighborhoods when performance is close to linear there. If the gradient is nonzero on the neighborhood, then performance converges to a linear model on small enough neighborhoods. If the gradient is zero, then higher order terms dominate on small neighborhoods.

For example, suppose that $\nabla f(z)$ vanishes at the centroid and the distribution is normal. Then, to quadratic approximation:
\begin{equation} \label{eqn: epsilon gaussian at peak}
\epsilon = \frac{H_{xy}^{\alpha \beta} H_{xy}^{\mu \nu} \Sigma_{\alpha \mu} \Sigma_{\beta \nu}}{ H_{xx}^{\alpha \beta} H_{xx}^{\mu \nu} \Sigma_{\alpha \mu} \Sigma_{\beta \nu} } = \frac{\langle H_{xy} \Sigma , \Sigma H_{xy} \rangle}{\langle H_{xx} \Sigma , \Sigma H_{xx} \rangle}
\end{equation}
where $\langle A, B \rangle = \sum_{i,j} a_{ij} b_{ij}$ denotes the matrix innter product.

Equation \ref{eqn: epsilon gaussian at peak} inspires interpretation. The numerator and denominator both consist of a tensor product between blocks of the Hessian and the trait covariance. The form of the product is the same in the numerator and denominator, only the block changes. The numerator depends on the off-diagonal skew symmetric block $H_{xy}$, while the denominator depends on the diagonal symmetric block $H_{xx}$. Thus, when the gradient vanishes, $\epsilon$ compares the size of these two blocks. 

The comparison becomes explicit if we work in a coordinate system where the traits, after evolution, are independent and all share the same variance. Then $\Sigma \propto I$. Such a change of coordinates exists whenever $\Sigma$ is full rank, that is, as long as the traits are not constrained to a lower dimensional subspace. Then, in this whitened coordinate system $\Sigma = \sigma^2 I$ for some $\sigma^2$ so:
\begin{equation}
    A^{\alpha \beta} A_{\mu \nu} \Sigma_{\alpha \mu} \Sigma_{\beta \nu} = \sigma^4 \sum_{i,j,k,l} a_{i,j} a_{k,l} \delta_{i,k} \delta_{j,l} = \sigma^4 \sum_{i,j} a_{i,j}^2 = \sigma^4 ||A||_{\text{Fro}}^2
\end{equation}
where $\delta_{i,j}$ is the discrete delta function, and $||A||_{\text{Fro}}$ denotes the Frobenius norm of the matrix $A$. Let $\tilde{H}$ denote the Hessian in the coordinate system where the traits are independent and normal. Then:
\begin{equation} \label{eqn: ratio of norms}
    \epsilon = \left( \frac{||\tilde{H}_{xy}||_{\text{Fro}}}{||\tilde{H}_{xx}||_{\text{Fro}}} \right)^2.
\end{equation}

Therefore, the factor $\epsilon$ is the ratio of the Frobenius norm of the off-diagonal block of the Hessian to the Frobenius norm of the diagonal block of the Hessian squared in the whitened coordinates. Competition is highly transitive when $||\tilde{H}_{xx}||_{\text{Fro}} \gg ||\tilde{H}_{xy}||_{\text{Fro}}$, and is highly cyclic when $||\tilde{H}_{xx}||_{\text{Fro}} \ll ||\tilde{H}_{xy}||_{\text{Fro}}$.

If $\nabla_x f(z,z) \neq 0$ then Equation \ref{eqn: ratio of norms} is modified by adding in a gradient dependent term to the denominator. The gradient term has the form $\nabla_x f(z,z)^{\intercal} \Sigma \nabla_x f(z,z)$ which equals $\sigma^2 ||\nabla_x \tilde{f}(z,z)||^2$ in the whitened coordinates. So, in the whitened coordinate system:
\begin{equation} \label{eqn: epsilon gaussian identity cov}
    \epsilon = \frac{||\tilde{H}_{xy}||_{\text{Fro}}^2}{2 \sigma^{-2} ||\nabla_x \tilde{f}(z,z)||^2 + ||\tilde{H}_{xx}||_{\text{Fro}}^2}.
\end{equation}

The larger the trait variance the more the terms associated with the Hessian dominate, while the smaller the variance the more the term associated with the gradient dominates, and the smaller $\epsilon$. For sufficiently small $\sigma$ the quantity $\epsilon$ is $\mathcal{O}(\sigma^2)$, so competition will approach perfect transitivity. We revisit this idea in a more general setting in the next section. 

As a final special case, suppose that the trait distribution converges to a Boltzmann type steady state with respect to expected performance. That is, $\pi_x(x) \propto \exp(\mathbb{E}_Y[f(x,Y)])$. Then, to quadratic approximation, $X \sim \mathcal{N}(z,\Sigma)$ where $\Sigma \propto H_{xx}^{-1}$. Boltzmann type models based on expected payouts are widely used for exploration in reinforcement learning and multi-armed bandit problems \cite{cesa2017boltzmann,schlag1998imitate}, in logistic fictitious play \cite{fudenberg1998theory}, and arise naturally in some population genetics models (c.f.~\cite{de2011contribution,sella2005application}). Distributions of this kind also arise naturally as solutions to the continuous trait replicator equation when initialized from a normal distribution on a neighborhood where performance is near to quadratic and where there is an internal maximum in expected performance. Cressman, Hofbauer and Riedel provide examples in \cite{Cressman_c}.

Now $\epsilon$ simplifies dramatically. If $X \sim \mathcal{N}(z,\Sigma)$ where $\nabla_x f(z,z) = 0$ and $\Sigma \propto H_{xx}(z,z)^{-1}$ then (see Appendix \ref{app: supplemental proofs}):
\begin{equation} \label{eqn: epsilon Boltsmann}
    \epsilon = \frac{1}{T} H_{xy}^{\alpha \beta} H_{xy}^{\mu \nu} {H_{xx}^{-1}}_{\alpha \mu} {H_{xx}^{-1}}_{\beta \nu} = \frac{1}{T} \langle H_{xy} H_{xx}^{-1}, H_{xx}^{-1} H_{xy} \rangle.
\end{equation}

While these expressions for $\epsilon$ depend on particular distributional assumptions and a quadratic performance function, they provide helpful intuition for how $\epsilon$, and, as a consequence, $\mathbb{E}[\|F_c\|^2]$, depend on $H_{xx}$, $H_{yy}$, $\nabla_x f(z,z)$, and the covariance $\Sigma$. In effect, the cyclic component of competition is only large if $\sigma^2 H_{xy}$ is large compared to $\nabla_x f(z,z)$, or $\nabla_x f(z.z)$ is small and $H_{xy}$ is large relative to $H_{xx}$. Otherwise, $\epsilon$ is small.
%In effect $\epsilon$ compares the size of $H_{xx}$ and $H_{yy}$, and is large when $H_{xy}$ is large relative to $H_{xx}$. If the distribution is not centered where a gradient in performance vanishes, then $\epsilon$ compares $H_{xy}$ to both the gradient and $H_{xx}$, though the gradient term will dominate if the covariance is small relative to the curvature in performance. 

\subsection{Trait Concentration} \label{sec: trait concentration}

What if performance is not quadratic, but the distribution of traits is concentrated on a small local neighborhood? How do the expected sizes of cyclic and transitive competition scale as the distribution concentrates?

Here we demonstrate that, in the limit as the trait distribution approaches a delta distribution, competition converges to perfect transitivity, and the rate of convergence is controlled by the quadratic approximation to $\epsilon$. There are many ways in which a distribution can concentrate, so we avoid explicit distributional assumptions. Instead we introduce an abstract concentration parameter $\kappa$ and consider families of trait distributions $\pi_x(\kappa)$ that converge to a delta distribution as $\kappa$ converges to zero. We consider different methods of controlling the rate of concentration in $\kappa$. At strongest, $\kappa$ is an explicit parameter in the distribution (say, the variance in a normal distribution). More weakly, $\kappa$ could bound the rate of convergence of the central moments to zero, or the rate at which a ball containing most of the probability mass collapses onto the centroid $z$. 

Concentration could also be defined by introducing a distance measure between distributions, then studying convergence to a delta distribution in the associated topology. This is the approach commonly adopted to prove the stability of monomorphic populations (delta distributions) in evolutionary game theory. We avoid this approach since different notions of distance induce different topologies (strong, or weak), and stability statements may depend on the chosen topology \cite{Cressman_b,cressman2004dynamic,oechssler2001evolutionary}. Future work could attempt the same proofs using convergence in a chosen topology.

A sample concentration parameterization follows. Consider a trait distribution with centroid $z$ and density function $\pi_x(x|\kappa)$ such that $\pi_x(x|\kappa) \propto \pi_x(y|1)$ where $y = z + (x - z)/\kappa$. Then, as $\kappa$ shrinks, the distribution retains the same shape, while concentrating about its centroid $z$. Here concentration is performed by contracting the distribution about its centroid while maintaining its form. Normal distributions provide a natural example that we will use running forward. Given $X \sim \mathcal{N}(z,\kappa^2 \Sigma)$, the associated trait distribution satisfies $\pi_x(x|\kappa) \propto \exp(-(x-z)^{\intercal}(\kappa^2 \Sigma)^{-1} (x - z)) = \exp(-\kappa^{-1}(x-z)^{\intercal}( \Sigma)^{-1} \kappa^{-1}(x - z)) \propto \pi_x(z + (x -z)/\kappa| 1)$.\footnote{An example mechanism: if evolution is governed by the replicator dynamic, the population is normally distributed, competition is quadratic, the gradient vanishes at $x_*$, and the Hessian $H_{xx}$ is negative definite, then $X \sim \mathcal{N}(\bar{x}(t),\kappa^2 \Sigma(t))$ where $\bar{x}(t) - x_* \rightarrow 0$ and $\Sigma(t) \rightarrow 0$ at rate $\mathcal{O}(t^{-1})$ \cite{Cressman_c,cressman2004dynamic}.} This parameterization makes the most sense for unimodal distributions. We refer to sequences of distributions that concentrate in this manner as spatially contracting distributions. When a distribution contracts about $z$, any nonzero central moment of degree $n$ will vanish proportional to $\kappa^n$. 

%Alternatively, we can define a family of concentrating distributions by raising a probability density function to increasing powers. Suppose that $\pi_x(x|\kappa) \propto \pi_x(x|1)^{1/\kappa}$. Then, writing $\pi_x(x|1) \propto \exp(-U(x))$ we have $\pi_x(x|\kappa) \propto \exp(-\kappa^{-1} U(x))$ which converges towards a series of delta functions at the modes of the density function. If the density function has a unique mode then $\pi_x(x|\kappa)$ converges to a delta function at the mode. Note the similarity here to varying temperature in a thermodynamic system governed by a potential $U(x)$. These sorts of limits are considered in simulated annealing and other methods based on Boltzmann distributions. The normal distribution also works here, provided we raise the original distribution to $\kappa^{-2}$.

%In either case, as $\kappa$ goes to zero the trait distribution collapses to a delta distribution. 
Note that, results from sequences of contracting distributions extend to specific fixed distributions  provided the fixed distribution can be treated as a member of a contracting sequence far enough in the tail for the limiting arguments to apply. In what follows we do not assume that the sequence of distributions contracts spatially, but impose bounds on the rates of convergence of specific moments, or tail probabilities instead.

In a concentration limit all of the terms in $\rho$ and $\epsilon$ converge to zero. So, to analyze the limits we must compare the rates of convergence. Suppose $\kappa$ is a scalar quantity that converges to zero. Suppose that $g(\kappa)$ and $h(\kappa)$ are both functions of $\kappa$ where $h(\kappa)$ is scalar valued and converges to zero as $\kappa$ converges to zero. Then $g(\kappa)$ is $\mathcal{O}_{\leq}(h(\kappa))$ if $\lim_{\kappa \rightarrow 0} g(\kappa)/h(\kappa) < \infty$. In that case $g(\kappa)$ goes to zero at least as fast as $h(\kappa)$. Alternately $g(\kappa)$ is $\mathcal{O}_{<}(h(\kappa))$ if $\lim_{\kappa \rightarrow 0} g(\kappa)/h(\kappa) = 0$. In that case $g(\kappa)$ goes to zero faster than $h(\kappa)$. In both cases $h(\kappa)$ sets an upper bound on the rate at which $g(\kappa)$ goes to zero. We will use upper bounds on rates of convergence to drop complicating terms that become negligible in the limit.

To drop the complicating terms, we also need lower bounds on the rates of convergence of the terms we wish to keep. To this end we say that a scalar valued function $g(\kappa)$ is $\mathcal{O}_{=}(h(\kappa))$ if:
\begin{equation}
   \lim_{\kappa \rightarrow 0} \frac{g(\kappa)}{h(\kappa)} = L \text{ where } 0 < |L| < \infty.
\end{equation}
Then $g(\kappa)$ goes to zero at the same rate as $h(\kappa)$. 

We will extend these definitions to describe the convergence rates of matrices and tensors storing central moments. A matrix or tensor is $\mathcal{O}_{<}(h(\kappa))$ if all of its entries are $\mathcal{O}_{<}(h(\kappa))$. For example, the tensor of fourth order moments is $\mathcal{O}_{<}(\kappa^2)$ if all of the fourth order central moments converge to zero faster than $\kappa^2$. 

When using $\mathcal{O}_{=}$ to establish equality in rates we need more specificity regarding zero. We say that a square-matrix valued function of $\kappa$, $G(\kappa)$, is $\mathcal{O}_{=}(\kappa)$ if all of its singular values are $\mathcal{O}_{=}(\kappa)$. Let $\sigma(G(\kappa))$ denote the singular values of $G(\kappa)$. Then $G(\kappa)$ is $\mathcal{O}_{=}(\kappa)$ if:
\begin{equation}
    \lim_{\kappa \rightarrow 0} \frac{G(\kappa)}{h(\kappa)} = G \text{ where } 0 < \sigma_{\text{min}}(G) \leq \sigma_{\text{max}}(G) < \infty.
\end{equation}

\subsubsection{Concentration for Quadratic Performance} \label{sec: concentration quadratic}

We begin by studying quadratic functions. We show that competition becomes increasingly transitive as $\kappa$ goes to zero when performance is quadratic, the covariance is proportional to $\kappa^2$, and the higher order moments vanish at a faster rate.
Then:

\vspace{2mm}

\textbf{Lemma 3: (Trait Concentration and $\epsilon$ for Quadratic Performance)} \textit{If $f$ is quadratic, the trait distribution $\pi_x(\kappa)$ depends on a concentration parameter $\kappa$, has mean $z$, positive definite covariance $\mathcal{O}_{=}(\kappa^2)$, third and fourth order central moments of $\mathcal{O}_{<}(\kappa^2)$, and $\nabla_x f(z,z) \neq 0$ then:
\begin{equation}
    \epsilon = \mathcal{O}_{\leq}(\kappa^2).
\end{equation}
and:
\begin{equation}
    \frac{\mathbb{E}[||F_c||^2]}{\mathbb{E}[||F||^2]} = \mathcal{O}_{\leq}(\kappa^2)
\end{equation}
with equality if and only if $H_{xy} \neq 0$. 
}

 \vspace{2mm}
 
The proof is provided in Appendix \ref{app: supplemental proofs}. The assumptions of Lemma 3 are automatically satisfied for any normal trait distribution with vanishing variance, or for any sequence of distributions  contracting spatially about $z$, provided the gradient is nonzero at $z$. Note that, setting the covariance to $\mathcal{O}_{=}(\kappa^2)$, assumes that the smallest and largest possible standard deviation along any direction in trait space are $\mathcal{O}_{=}(\kappa)$.

It follows that, if performance is quadratic, and the trait distribution concentrates where $\nabla_x f(z,z) \neq 0$, then competition converges to perfect transitivity and the relative size of the cyclic component converges to zero at least as fast as the variance in the trait distribution. %Note that if traits do not interact, or the trait space is one dimensional, then $H_{xy} = 0$ automatically, so we expect unusually fast convergence to transitivity.

In fact, all components of competition will vanish in the limit since similar competitors must be close to evenly matched when $f$ is continuous. The absolute sizes of the components of competition are controlled by the variance in performance, $\mathbb{V}_{X,Y}[f(X,Y)]$. By tracking the rate at which the variance converges to zero, and applying the trait-performance theorem, we recover the rates at which each component vanishes.

\vspace{2mm}
\textbf{Lemma 4: (Vanishing Variance for Quadratic Performance)} \textit{If $f$ is quadratic, the trait distribution $\pi_x(\kappa)$ depends on a concentration parameter $\kappa$, has mean $z$, positive definite covariance $\mathcal{O}_{=}(\kappa^2)$, third and fourth order central moments of $\mathcal{O}_{<}(\kappa^2)$, and $\nabla_x f(z,z) \neq 0$, then:
\begin{equation}
    \mathbb{V}_{X,Y}[f(X,Y)] = \mathcal{O}_{\leq}(\kappa^2)
\end{equation}
and:
\begin{equation} \label{eqn: component convergence rates}
\begin{aligned}
    & \mathbb{E}[||F||^2] = \mathcal{O}_{=}(\kappa^2) \\
    & \mathbb{E}[||F_t||^2] = \mathcal{O}_{=}(\kappa^2)\\
    & \mathbb{E}[||F_c||^2] = \mathcal{O}_{\leq}(\kappa^4)
\end{aligned}
\end{equation}
with equality if and only if $H_{xy} \neq 0$. 
}
\vspace{2mm}

See Appendix \ref{app: supplemental proofs} for the proof, which follows closely from the proof of Lemma 3. Note that, while all of the components converge to zero in the concentration limit, the cyclic component vanishes faster than the rest, producing ever more transitive tournaments.  

What if the distribution concentrates at some $z$ where the gradient vanishes? Then:
$$
\begin{aligned}
\epsilon = \frac{1}{2}\frac{H_{xy}^{\alpha \beta} H_{xy}^{\mu \nu} \text{Cov}[X]_{\alpha \mu} \text{Cov}[X]_{\beta \nu}}{ \frac{1}{4} H_{xx}^{\alpha \beta} H_{xx}^{\mu \nu} \left(\text{Fo}[X]_{\alpha \beta \mu \nu} - \text{Cov}[X]_{\alpha \beta} \text{Cov}[X]_{\mu \nu}  \right) } . 
\end{aligned}
$$

The denominator is the variance in the local rating function when the gradient is zero, so is nonnegative. It follows that the fourth order central moments are $\mathcal{O}_{=}(\kappa^4)$ whenever the trait covariance is  $\mathcal{O}_{=}(\kappa^4)$. Then the denominator and numerator have the same order in $\kappa$ so $\epsilon$ will not vanish as $\kappa$ goes to zero. For example, when the trait distribution is normal, $\epsilon$ is given by a ratio of matrix norms (see Equations \ref{eqn: epsilon gaussian at peak} and \ref{eqn: ratio of norms}). 

So, in the case when $\nabla_x f(z,z)$ vanishes:

\vspace{2mm}

\textbf{Lemma 5: (Trait Concentration when the Gradient Vanishes)} \textit{If $f$ is quadratic, the trait distribution $\pi_x(\kappa)$ depends on a concentration parameter $\kappa$, has mean $z$, positive definite covariance $\mathcal{O}_{=}(\kappa^2)$, third and fourth order central moments of $\mathcal{O}_{<}(\kappa^2)$, but $\nabla_x f(z,z) = 0$, then:
\begin{equation}
    \epsilon = \mathcal{O}_{\leq}(1)
\end{equation}
with equality if and only if $H_{xy} = 0$, and:
\begin{equation}
    \mathbb{V}_{X,Y}[f(X,Y)] = \mathcal{O}_{\leq}(\kappa^4)
\end{equation}
with equality if and only if $H_{xx} \neq 0$. If $H_{xx} \neq 0$ then:
\begin{equation}
    \begin{aligned}
    & \mathbb{E}[||F||^2] = \mathcal{O}_{=}(\kappa^4) \\
    & \mathbb{E}[||F_t||^2] = \mathcal{O}_{=}(\kappa^4) \\
    & \mathbb{E}[||F_c||^2] = \mathcal{O}_{\leq}(\kappa^4) \\   
\end{aligned}
\end{equation} }
\textit{with equality if and only if $H_{xy} \neq 0$.}

\vspace{2mm}

The proof follows exactly from the arguments used before, only without the lowest order gradient term in the denominator of the expression for $\epsilon$. Therefore, when the gradient vanishes at the centroid, the expected relative size of the cyclic component can converge to a nonzero value. Specifically:
\begin{equation}
    \lim_{\kappa \rightarrow 0} \frac{\mathbb{E}[\|F_c\|^2]}{\mathbb{E}[\|F\|^2]} = \frac{\epsilon}{1 + \epsilon} \frac{L}{E} \text{ where }  \lim_{\kappa \rightarrow 0} \epsilon = 2 \frac{H^{\alpha \beta}_{xy} H^{\mu \nu}_{xy} \Sigma_{\alpha \mu} \Sigma_{\beta \nu}}{H^{\alpha \beta}_{xx} H^{\mu \nu}_{xx}(\text{Fo}[X]_{\alpha \beta \mu \nu} - \Sigma_{\alpha \mu} \Sigma_{\beta \nu})}
\end{equation}
where $\Sigma$ is the limiting trait covariance (scaled by $\kappa^{-2}$) and $\text{Fo}[X]$ is the limiting tensor of central fourth order moments (scaled by $\kappa^{-4}$). Here we retain a cyclic component in the concentration limit since the local linear model is equal to zero. 

Combined, Lemmas 3, 4, and 5 fully characterize how the expected sizes of cyclic and transitive competition behave for quadratic performance functions, provided the Hessian blocks are nonzero. If the gradient and the Hessian blocks are zero then we are forced to look at higher order terms in the expansions of the numerator and denominator. These depend on ever higher order moments in $\kappa$. Figure \ref{fig:Convergence rate flowchart} illustrates the sequence of convergence rates given the lowest order nonzero terms in the numerator and denominator of $\epsilon$.

Generically, the gradient or Hessian blocks do not vanish, so Lemma 3 acts as the general case. However, there are good reasons why the gradient or off diagonal block of the Hessian may vanish. The off-diagonal block of the Hessian vanishes when traits are non-interacting, or there is only one trait, as outlined in Lemma 1. The gradient may vanish if the trait distribution is the steady state of an evolutionary process that concentrates at some point in the interior of the state space. If the gradient is nonzero, a concentrated distribution should move in the direction of the gradient. Thus, it is plausible that concentration on the interior of the trait space should occur where the gradient is small, if not zero. In contrast, a distribution may concentrate under the pressure of a nonzero gradient against the boundary of the trait space. 

\subsubsection{Concentration for General Performance} \label{sec: concentration generic}

General performance functions are not quadratic. Yet, if a performance function is smooth, then it may be approximated locally with a quadratic function. As a distribution concentrates it focuses on a small neighborhood, so the quadratic expansion ought to predict the limiting behavior, provided the associated errors vanish fast enough. 

The correlation $\rho$ is bounded by (see Equation \ref{eqn: epsilon}, \ref{eqn: epsilon bound on rho}):
$$ 
\rho \leq \frac{1}{2} \frac{1}{1 + \epsilon} \text{ where } \epsilon = \frac{\mathbb{E}_{X,Y}[h(X,Y|z)^2]}{2 \mathbb{V}_{X}[r(X|z)] + 4 \mathbb{E}_{X,Y}[r(X|z) h(X,Y|z)]}
$$
for any choice of local rating function that averages to zero, and $z$ set to $\mathbb{E}[X]$. As usual we will use the quadratic local rating function. When considering non-quadratic performance functions the bound may not be tight, and $h(x,y|z)$ will include more terms than the quadratic block involving $H_{xy}$. So, write:
\begin{equation}
    h(x,y|z) = (x - z)^{\intercal} H_{xy}(z,z) (x - z) + g(x,y|z)
\end{equation}
where $g(x,y|z)$ is the error between $f(x,y)$ and its local quadratic approximation at $z,z$. We already showed that $\mathbb{E}_{X,Y}[r(X|z) (x - z)^{\intercal} H_{xy} (x - y)] = 0$ so $\epsilon$ can be written:
$$
\epsilon = \frac{\mathbb{E}_{X,Y}[\left((X - z)^{\intercal} H_{xy}(z,z) (Y - z) \right)^2] + 2 \mathbb{E}[(X - z)^{\intercal} H_{xy}(z,z) (Y - z) g(X,Y|z)] + \mathbb{E}_{X,Y}[g(X,Y|z)^2]}{2 \mathbb{V}_{X}[r(X|z)] + 4 \mathbb{E}_{X,Y}[r(X|z) g(X,Y|z)]}
$$

The numerator and denominator differ from the quadratic case by the expectations involving $g$. In what follows, we show that these errors usually vanish fast enough that $\epsilon$ converges to its quadratic approximation. 

That argument is developed in a sequence of lemmas. It separates the error in into two components, a local component that can be bounded by the central moments, and a tail component that can be bounded by ensuring that the tails of the distribution vanish quickly. The first lemma establishes the necessary moment scaling needed to drop the error terms. The second shows that, if the support of the distribution collapses to zero, then, the moments will collapse at the rates required for convergence to the quadratic approximation. The last establishes that, if the support does not collapse to zero, but the tails of the distribution vanish quickly enough, then the distribution can be approximated with a windowed distribution whose support collapses to zero.

The proofs are provided in the Appendix \ref{app: supplemental proofs}, so we sketch the main arguments here before providing the technical statements. All of the statements depend on a smoothness assumption on $f$, which ensures that local quadratic approximation is possible, and a concentration assumption which controls the limiting behavior of $\pi_x$ and its moments.

\vspace{2mm}
\textbf{Lemma 6: (Power Series $f$ and Vanishing Moments)} \textit{Suppose that:}

\begin{enumerate}
\item \textit{$f(x,y)$ admits a globally convergent Taylor expansion about $x = y = z$ for all $z \in \Omega$, and,}
\item \textit{$\pi_x$ is a trait distribution with centroid $z$ and concentration parameter $\kappa$ such that the trait covariance is $\mathcal{O}_{=}(\kappa^2)$, higher order central moments of degree $n \leq 5$ are order $\mathcal{O}_{\leq}(\kappa^n)$, and all other higher order central moments are order $\mathcal{O}_{<}{\kappa^5}$.} 
\end{enumerate}

\textit{Then, provided $\nabla_x f(z,z) \neq 0$ or $H_{xx}(z) \neq 0$, $\epsilon$ and $\mathbb{V}_{X,Y}[f(X,Y)]$ converge to their approximations using the local quadratic model. Moreover, convergence to $\epsilon$ occurs at least one order faster than $\epsilon$ converges to zero.} 

\vspace{2mm}

The moment convergence rates introduced here ensure that errors arising from higher order terms converge to zero faster than matching terms in the quadratic approximation. We require that moments of order $n \in [2,3,4,5]$ all vanish at rates less than or equal to $n$ so that the higher order terms all vanish at least one order faster than the terms we wish to keep, which at the fastest, converge to zero at $\mathcal{O}_{=}(\kappa^4)$. The central moments will converge to zero at these rates if concentration is governed via spatial contraction about $z$. For example, if $X \sim \mathcal{N}(z, \kappa^2 \Sigma)$, then $\pi_x(\kappa x|\kappa) \propto \pi_x(x|1)$, so the $n^{th}$ order moments are all $\mathcal{O}_{\leq}(\kappa^n)$ with equality if the $n^{th}$ order moment is nonzero at $\kappa = 1$. 

The following lemmas do not assume that concentration occurs via spatial contraction, or enforce specific moment scaling, but introduce a collapsing ball about $z$ which ensures that the higher order moments vanish fast enough to satisfy the moment bounds used here. The following lemma also relaxes the smoothness assumption on $f$.

\vspace{2mm}
\textbf{Lemma 7: (Quadratically Approximable $f$ and Collapsing Support)} \textit{Suppose that:}

\begin{enumerate}
\item \textit{for all $z \in \Omega$, $f(x,y)$ is second differentiable at all $x = y = z$ and the second order Taylor expansion of $f(x,y)$ about $x = y = z$ has errors that, on some ball centered at $z$ with radius $r(z) > 0$, are bounded by a power series of $(x - z)$ and $(y-z)$ whose lowest order terms are cubic, and }

\item \textit{$\pi_x$ is a trait distribution with centroid $z$ and concentration parameter $\kappa$ such that such that the trait covariance $\text{Cov}_{X\sim \pi_x}[X]$ is $\mathcal{O}_{=}(\kappa^2)$. Suppose in addition that there exists a ball $B_{R(\kappa}(z)$ centered at $z$ with radius $R(\kappa)$ that covers the support of the trait distribution and where $R(\kappa)$ converges to zero as $\kappa$ goes to zero at order $\mathcal{O}_{<}(\kappa^{4/5})$.}
\end{enumerate}

\textit{Then, provided $\nabla_x f(z,z)\neq 0$ or $H_{xx}(z) \neq 0$, $\epsilon$ and $\mathbb{V}_{X,Y}[f(X,Y)]$ converge to their local quadratic approximations, and convergence to $\epsilon$ occurs faster than $\epsilon$ converges to zero.}   

\vspace{2mm}

The proof follows as, if the support is contained inside of a collapsing ball, then it is eventually contained inside of the ball where the errors in the quadratic approximation are bounded by a power series whose lowest order terms are cubic. The resulting errors can then be expressed in terms of the central moments of the trait distribution. Those moments vanish in the concentration limit at rates bounded by the collapse of the support. The rate at which they vanish is controlled by the rate at which $R(\kappa)$ vanishes, hence the careful choice of the convergence rate of $R(\kappa)$. As long as $R(\kappa)$ vanishes faster than $\kappa^{4/5}$, the third order moments vanish faster than $\kappa^2$, the fourth order moments vanish faster than $\kappa^3$, and fifth and higher order moments vanish faster than $\kappa^4$, ensuring convergence to the quadratic approximand. We do not assume $R(\kappa) = \mathcal{O}_{=}(\kappa)$, as, in the subsequent analysis, we aim to take $R(\kappa)$ to zero slower than $\kappa$.

What if the support of the trait distribution does not collapse onto $z$? The concentration result still holds, provided there is a ball collapsing about $z$ that contains most of the probability mass. We need one more technical lemma to ensure that errors contributed by the tails of the trait distribution can be ignored. 

Let $B_{R(\kappa)}(z)$ denote a ball centered at $z$ with radius $R(\kappa)$ where $R(\kappa) \rightarrow 0$ as $\kappa \rightarrow 0$. Let $p(\kappa)$ denote the probability of sampling $X$ outside the ball. Define the windowed distribution $\pi_x^w$:
\begin{equation}
    \pi_x^w(x|\kappa) = \frac{1}{1 - p(\kappa)} \chi_{B_{R(\kappa)}(z)}(x) \pi_x(x|\kappa)
\end{equation}
where $\chi_{S}(x)$ is the indicator function for the set $S$. Then $\pi_w$ is the distribution of $X$ conditioned on drawing $X$ inside the ball. We aim to replace $\epsilon$ and $\mathbb{V}_{X,Y}[f(X,Y)]$ with their windowed approximations given by replacing $\pi_x$ with $\pi_x^w$. These windowed approximations converge provided $p(\kappa)$ goes to zero fast enough. The following lemma establishes conditions under which the windowed approximation of a generic observable (expectation of some function of $x$ and $y$) converges with rate $\mathcal{O}_{\leq}(p(\kappa))$. 

\vspace{2mm}
\textbf{Lemma 8: (Negligible Tails)} \textit{Suppose that: }
\begin{enumerate}
\item \textit{$g(x,y)$ is an arbitrary, function of $x,y$ that is bounded in magnitude on $\Omega \times \Omega$, and}
\item \textit{$\pi_x$ is a trait distribution with centroid $z$ and concentration parameter $\kappa$ such that such that there exists a ball $B_{R(\kappa}(z)$ centered at $z$ with radius $R(\kappa)$ which converges to zero as $\kappa$ goes to zero, and where the probability $p(\kappa)$ of sampling $X$ outside the ball converges to zero as $\kappa$ goes to zero.}
\end{enumerate}

\textit{Then $\mathbb{E}_{X,Y \sim \pi_x}[g(X,Y)]$ converges to its windowed approximation  $\mathbb{E}_{X,Y \sim \pi_x^w}[g(X,Y)]$ at order $\mathcal{O}_{\leq}(p(\kappa))$.} 

\vspace{2mm}

Then, combining our lemmas: 

\vspace{2mm}
\textbf{Theorem 2: (Trait Concentration)} \textit{Suppose that $f(x,y)$ is a bounded performance function on $\Omega \times \Omega$, and satisfies the smoothness assumptions of Lemma 7. Suppose that $\pi_x$ is a trait distribution that satisfies the concentration assumptions of Lemma 8. If there exists a $R(\kappa)$ which converges to zero at rate $\mathcal{O}_{<}(\kappa^{4/5})$ while $p(\kappa)$ converges to zero at rate $\mathcal{O}_{<}(\kappa^4)$, then $\epsilon$ and $\mathbb{V}_{X,Y}[f(X,Y)]$ converge to their approximations using the local quadratic model of $f(x,y)$ about $x = y = z$ with errors vanishing faster than the approximations specified by Lemmas 2 and 3. Then the expected sizes of the components of competition are governed by Lemmas 3 and 4 so:
$$    
\begin{aligned}
    & \mathbb{E}[||F||^2] = \mathcal{O}_{=}(\kappa^2) \\
    & \mathbb{E}[||F_t||^2] = \mathcal{O}_{=}(\kappa^2)\\
    & \mathbb{E}[||F_c||^2] = \mathcal{O}_{\leq}(\kappa^4)
\end{aligned}
$$
if $\nabla_x f(x,y)|_{x=y=z} \neq 0$ and with equality if and only if $H_{xy}(z,z) \neq 0$. If $\nabla_x f(x,y)|_{x=y=z} = 0$  and $H_{xx}(z,z) \neq 0$ then:
$$
\begin{aligned}
    & \mathbb{E}[||F||^2] = \mathcal{O}_{=}(\kappa^4) \\
    & \mathbb{E}[||F_t||^2] = \mathcal{O}_{=}(\kappa^4) \\
    & \mathbb{E}[||F_c||^2] = \mathcal{O}_{\leq}(\kappa^4) \\
\end{aligned}
$$
with equality if and only if $H_{xy}(z,z) \neq 0$. }

\vspace{2mm}

Typically $\nabla_x f(x,y)|_{x=y=z}$ will only equal zero on a set of measure zero in $\Omega$. Thus, provided $f$ is sufficiently smooth, and the trait distribution concentrates sufficiently quickly, similarity will suppress cyclicity almost everywhere in $\Omega$. 

A last technical note: the theorem statement requires the existence of a ball with radius $R(\kappa) = \mathcal{O}_{<}(\kappa^{4/5})$ such that the probability $p(\kappa) = \text{Pr}\{X \notin B_{R(\kappa)}(z) \}$ is order $\mathcal{O}_{<}(\kappa^4)$. This is possible for most distributions with exponentially decaying tails that contract regularly towards zero as $\kappa$ goes to zero. For example, consider a one dimensional trait space with $\pi_x(x|\kappa) = \frac{1}{2}\kappa \exp(-|x|/\kappa)$. Then $p(\kappa)  = \exp(-R(\kappa)/\kappa)$ so setting $R(\kappa) = \kappa^{4/5}$ gives $p(\kappa) = \exp(-\kappa^{-1/5})$. Then $\lim_{\kappa \rightarrow 0} \kappa^n \exp(-\kappa^{-1/5}) = 0$ for any $n$. It follows that $p(\kappa)$ is $\mathcal{O}_{<}(\kappa^n)$ for any $n$. Similar results follow for the normal distribution, or other distributions with exponential tail decay.  

%%%%%%%%%%%%%%%%%%%%%%%%%%%%%%%%%%%%%%%%%%%%%%%%%%%%%%
\section{Numerical Demonstration} \label{sec: numerics}

To test our theory, we simulate a Gaussian adaptive process on a series of bimatrix games and random performance functions. We test whether evolution promotes concentration towards a small subset of the strategy space, and, consequently, promotes transitivity at the convergence rates predicted by Theorem 2.
\begin{table}[h!]
\begin{centering}
{\rowcolors{2}{white!80!}{gray!10!}
\begin{tabular}{ |p{4cm}| p{4cm}|p{4cm}|  }
 \hline
 \multicolumn{3}{|c|}{List of bimatrix games} \\
 \hline
 Games& Dilemma &Payout Matrix\\
 \hline
 Prisoner's dilemma &  Trust   & \begin{centering} $\begin{matrix}
(2,2) & (0,3)\\
(3,0) & (1,1)
\end{matrix}$ \end{centering}\\
 Stag hunt & Cooperation & $\begin{matrix}
(6,6) & (1,3)\\
(3,1) & (2,2)
\end{matrix}$\\
 Chicken   &  Escalation   & $\begin{matrix}
(1000,1000) & (999,1001)\\
(1001,999) & (0,0)
\end{matrix}$\\
 \hline
\end{tabular}}

 \end{centering}
 \vspace{2mm}
 \caption{List of bimatrix games considered and their payout matrices. Our Moran process simulation could only work with zero or positive values, so we converted all of the payouts to non-negative.}
\label{tab: bimatrix games list}
\end{table}

We first consider three canonical bimatrix games: chicken, the prisoner's dilemma, and stag hunt. Table \ref{tab: bimatrix games list} shows the payout matrices for each game.

Bimatrix games provide a simple, familiar, and have well-documented Nash equilibria. Taxonomies and reviews of such games can be found in \cite{bruns2015names,liebrand1983classification}. The prisoner's dilemma, gives each prisoner the choice to cooperate or defect, and is designed to model dilemmas involving trust and cooperation \cite{Skyrms}. Once iterated, it can model the evolution of altruism \cite{axelrod1981evolution}. Stag hunt also models cooperation \cite{Skyrms}. Each individual can independently hunt a hare for a small guaranteed payout, or can choose to hunt a stag for a higher payout. It takes both players to catch the stag, so a player who chooses to hunt the stag runs the risk that his partner chose to hunt a hare. The game of chicken has been used to models escalation problems, including nuclear conflict \cite{Rapoport}. It presents competitors with the choice to swerve or stay the course, if only one competitor stays they gain a small benefit, but if both stay they crash and die. 

Next, we convert the bimatrix payouts into a performance function. Performance could be defined by payout matrices, as when considering individual interactions. However, since our focus is on the population-level dynamics of the network, we consider a population-level payout instead. Note that, expected payout given a pair of mixed strategies is necessarily quadratic. A mixed distribution over two choices is parameterized by one degree of freedom. All quadratic games in one trait are perfectly transitive (see Lemma 1), so would not allow any exploration of convergence to transitivity. In contrast, population level processes allow nontrivial structure. 

For each bimatrix game, we determine event outcomes via a Moran process \cite{lieberman}. We initialize a population of individuals where half adopt strategy $A$ and half adopt $B$. Fitness is determined by game payouts. The process terminates at fixation, i.e.~when all individuals are of one type. The fixation probability given a pair of strategies, acts as a performance function. In this context the game is zero-sum; only one population can fix. Nevertheless, population-level processes can reward cooperation, even in a zero-sum setting, since cooperative agents receive large payouts in predominantly cooperative populations.

\begin{figure}[t]
    \centering
    \includegraphics[scale=0.5]{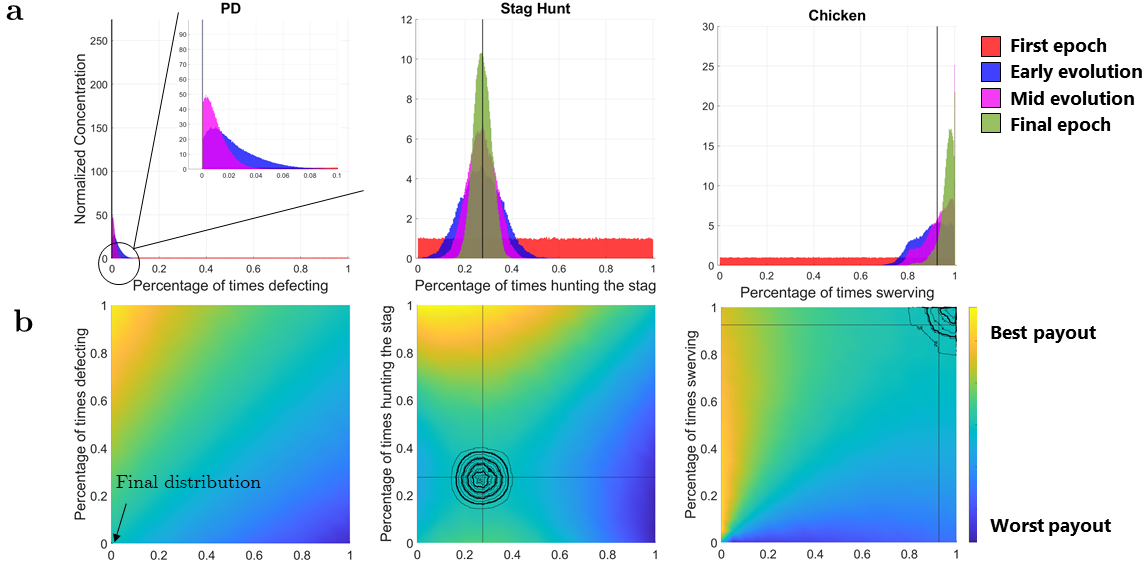}
    \caption{Trait concentration and bimatrix game performances. \textbf{(a)}: The normalized concentration of the strategies for different bimatrix games. The vertical black line is the predicted Nash equilibrium of our performance function. \textbf{(b)}: The performance heatmap for our Moran process. The contours indicate the final locations of the competitors, and the horizontal and vertical lines are the predicted Nash equilibrium.}
    %(blue: (epoch 3 for PD, 5 for stag and chicken)
    %(magenta: (epoch 5 for PD, 10 for stag and chicken}
    \label{fig:trait distribution and heatmaps}
\end{figure}

Running the Moran process to simulate every event outcome proved prohibitively expensive. Therefore, we approximated the fixation probabilities by interpolating sampled outcomes on a grid. We used the MATLAB curve fitting tool to generate a performance function that closely approximates the fixation probabilities. For the stag hunt and prisoner's dilemma games, cubic polynomials fit the data with no visible systematic errors (R-squared values $\geq$ 0.9993). We used the fits to compute the gradients and Hessian used throughout the convergence theory. For chicken, the extreme behavior at the corners of the strategy space (see the cusp in the bottom left hand corner of Figure \ref{fig:trait distribution and heatmaps}) created systematic errors so we used a cubic spline interpolant instead, and used numerical differentiation to approximate the gradient and Hessian.

Competitive outcomes between any pair of competitors were then determined using the performance function. Each competitor plays some number of games (usually 100) against a random selection of the other competitors drawn from the current population, and the 10 percent of competitors with the highest win rate are selected to reproduce. Children are assigned the same traits as their parent, plus a normal random vector scaled by a genetic drift parameter. If the sampled traits fall outside the trait space, they are projected back onto the boundary of the trait space instead. Every phase of competition, selection, and reproduction is one ``epoch".

At every epoch in the process, we perform a Helmholtz-Hodge Decomposition (HHD) of the complete graph of competitors to calculate the size of the transitive and intransitive components. These are normalized to produce proportions of transitivity and cyclicity as defined in Section \ref{sec: definitions} and \cite{strang2020applications,strang2022network}. To evaluate concentration, we calculated the covariance of the traits and the number of clusters, counted using a Gaussian mixture model. We allowed evolution to proceed until epoch 50, or when we had one stable cluster, whichever came first.

As a sanity check, we compared the location of the final trait distributions, to the Nash equilibria for the bimatrix games. Prisoner's dilemma has two pure-strategy Nash equilibria: either both agents always cooperate or always defect. The always cooperate strategy, has a higher expected payout for both competitors, but is invasible by a defecting strategy. In stag hunt, there are two pure-strategy Nash equilibria along with a mixed-strategy to hunt the stag with probability 0.25. For chicken, there is a pure-strategy Nash equilibrium to always go straight, and a mixed-strategy equilibrium to swerve with probability 0.999. While these Nash equilibria provide approximate locations for evolution to settle, in practice, the payout governing evolution is directed by the fixation properties from the Moran process. These equilibria are close to, but not exactly, the equilibria observed for the bimatrix games. In general our processes converged towards stable distributions centered near the Nash equilibria of the Moran process (see Figure \ref{fig:trait distribution and heatmaps}).

In most experiments ran we observed convergence towards transitivity driven by increasing concentration. Our theory predicts the rate of convergence to transitivity in concentration. This rate depends on the smoothness of the performance function, as measured by the norms of its low order partial derivatives. We tested the accuracy of these predicted rates as follows. We used Equation \ref{eqn: trait performance} from Theorem 1 to estimate $\rho$ by comparing the average sizes of the transitive and cyclic components in randomly sampled ensembles about the final cluster centroid. We drew each ensemble from a normal distribution, centered at the final cluster centroid, while taking the covariance to zero. We compared our empirical estimate to our analytic prediction based on Equation \ref{eqn: epsilon gaussian identity cov} to confirm Lemma 3, Lemma 5, and Theorem 2.

While the bimatrix games provide a widely studied measure for competition, they all have a one-dimensional trait space. Games in one-dimensional spaces are a special, highly transitive case (see Lemma 1). Therefore, we also considered randomly-generated, $n$-dimensional performance functions with tuneable structure chosen to illustrate the generality of our theory.

We designed our performance functions to satisfy the following properties. First, we desired tuneable smoothness, so we constructed the function as a sparse sum of Fourier modes with variable amplitudes and frequency. By increasing the low order modes we promote transitivity over neighborhoods of a fixed size. Second, we ensured that distinct traits of distinct competitors interact to produce attribute tradeoffs (e.g.~ speed versus strength). Such tradeoffs are, generically, the source of cyclic competition arising at lowest order in the Taylor expansion of performance. In special cases when distinct traits do not interact, the degree of cyclic competition vanishes exceptionally fast during concentration. Lastly, we skew symmetrized our function to enforce fairness, that is, $f(x,y) = -f(y,x)$.

The resulting performance function took the form:
\begin{equation}
\begin{aligned}
    f(x,y|\mathcal{P},\alpha,\phi,m) = \sum_{k=1}^m \sum_{i,j \in \mathcal{P}(k)} \frac{\alpha_{i,j}(k)}{k^2}  & \left( \sin(2 \pi k (x(i) - \phi_{i,j}(k))) \cos(2 \pi k (y(j) - \phi_{i,j}(k))) - \hdots \right. \\
    & \left. \sin(2 \pi k (y(i) - \phi_{i,j}(k))) \cos(2 \pi k (x(j) - \phi_{i,j}(k))) \right)
\end{aligned}
\end{equation}
where $\mathcal{P}(k)$ is the collection of interacting trait pairs at frequency $k$, $\alpha$ are the amplitudes, $\phi$ are a collection of phase shifts, and $m$ is the max frequency (and number of distinct frequencies) considered. Note that taking a difference of the form $f(x) g(y) - f(y) g(x)$ is automatically alternating in $x$ and $y$. We divided the amplitudes at higher frequencies by $k^2$ so that each frequency contributes equally to the Hessian. 

Then, our evolution test continued in the same manner as before, initialized with a uniformly sampled population selected from an $n$-dimensional trait hypercube with sides [-1,1].%\cc{This would be a location where we could talk about the convergence tests if we ended up considering them.}

%For our final example, we considered a simplified version of the Colonel Blotto game. Under these parameters, we would repeat the $n$-dimensions of traits, with the stipulation that both competitors have the same force and must divide it across $n$ different battlefields with pre-selected reward values known to the competitors. The performance functions were sigmoidal functions applied to the difference in force allocation between the two competitors on the battlefield, added together to make a performance for one strategy against one another. Once there, we continued the evolution test in the same way as for bimatrix games and the random performance function models.

\subsection{Bimatrix Games} \label{sec: bimatrix}

Figure \ref{fig:trait distribution and heatmaps}, showed the locations of the strategies on the trait space at different steps across evolution under control parameters. In prisoner's dilemma, evolution promotes quick convergence to the pure-strategy Nash equilibrium to always cooperate, and by even the first steps of evolution, nearly all the competitors almost always cooperate (concentrate at 0). For stag hunt, the final population distribution has Gaussian structure centered about probability 0.27 to hunt the stag (the NE for the Moran process). The earlier stages are also nearly Gaussian centered at 0.27 with standard deviation decreasing over time. Thus, the trait distribution concentrates as time progresses. Likewise, chicken has a truncated Gaussian structure centered near the NE, with standard deviations decreasing over evolution. Here the distribution abuts the boundary so, projection onto the boundary produces a second mode at swerve probability 1.

\begin{figure}
    \centering
    \includegraphics[scale=0.55]{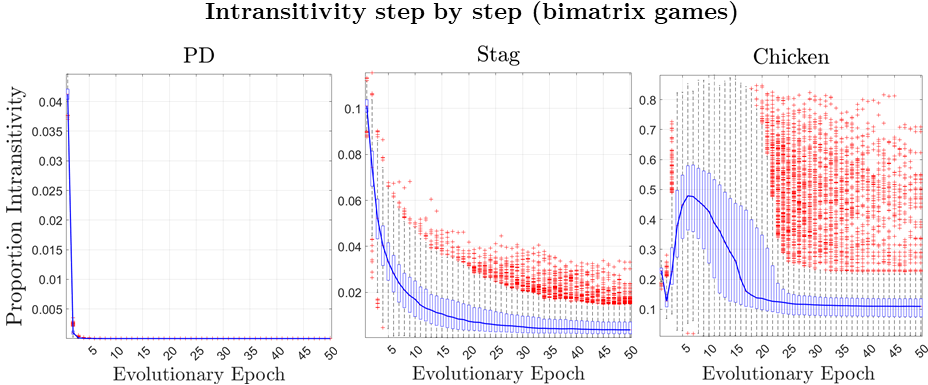}
    \caption{Step-by-step intransitivity for the three bimatrix games under control parameters. \textbf{(Left):} Prisoner's dilemma, \textbf{(Center):} Stag hunt, \textbf{(Right):} Chicken.}
    \label{fig: step by step bimatrix}
\end{figure}

Figure \ref{fig: step by step bimatrix} shows the proportion intransitivity per epoch observed in each bimatrix game, under control parameters. Note that the evolution of transitivity over time is different in each game. In the prisoner's dilemma example, the intransitivity immediately goes to zero. In the stag hunt example, the intransitivity decreases step by step, ending at less than 0.01 proportion. In chicken, however, the proportion intransitivity grows dramatically during the early stages of evolution, before settling close to 0.1. Thus, unlike the previous games, chicken sustains appreciable intransitivity.

Next, we varied the simulation parameters to test which parameters influence the observed results and how. Table \ref{table: bimatrix parameters} shows the control parameters chosen for the bimatrix game tests, along with individual parameter perturbations. Perturbation results were compared to results under the control scenario.
\begin{table}[h!]
\begin{centering}
\rowcolors{2}{white!80!}{gray!10!}
\begin{tabular}{ |p{4cm}| p{4cm}|p{4cm}|  }
 \hline
 \multicolumn{3}{|c|}{List of variables of consideration} \\
 \hline
 Parameter& Control & Perturbations \\
 \hline
 Fit (PD and stag) &   cubic  & quintic\\
 Interpolant (Chicken) & cubic spline & \\
 Genetic drift& $5 \cdot 10^{-3}$  & $5 \cdot 10^{-5}$, $1 \cdot 10^{-4}$, $5 \cdot 10^{-4}$, $1 \cdot 10^{-3}$, $1 \cdot 10^{-2}$, $5 \cdot 10^{-2}$\\
 Games played per epoch & 100  & 10, 1000\\
 \hline
\end{tabular}

\end{centering}
\vspace{4mm}
 \caption{List of parameter values and variations tested in the bimatrix games.}
 \label{table: bimatrix parameters}
\end{table}

Two of our parameters--the number of games played by each competitor per epoch and the type of interpolated performance function--did not influence the intransitivity over time significantly. 

The genetic drift parameter, in contrast, is highly significant. It determines how much the process explores the space, and how tightly it concentrates. Prisoner's dilemma is not sensitive to the genetic drift parameter since selection to the boundary is very strong. For stag hunt, higher values of genetic drift lead to an increase in the proportion of intransitivity, as the resulting population is less concentrated. Nevertheless, the general trend in intransitivity over time is consistent across all of the values of genetic drift tested. 

\begin{figure}
    \centering
    \includegraphics[scale=0.26]{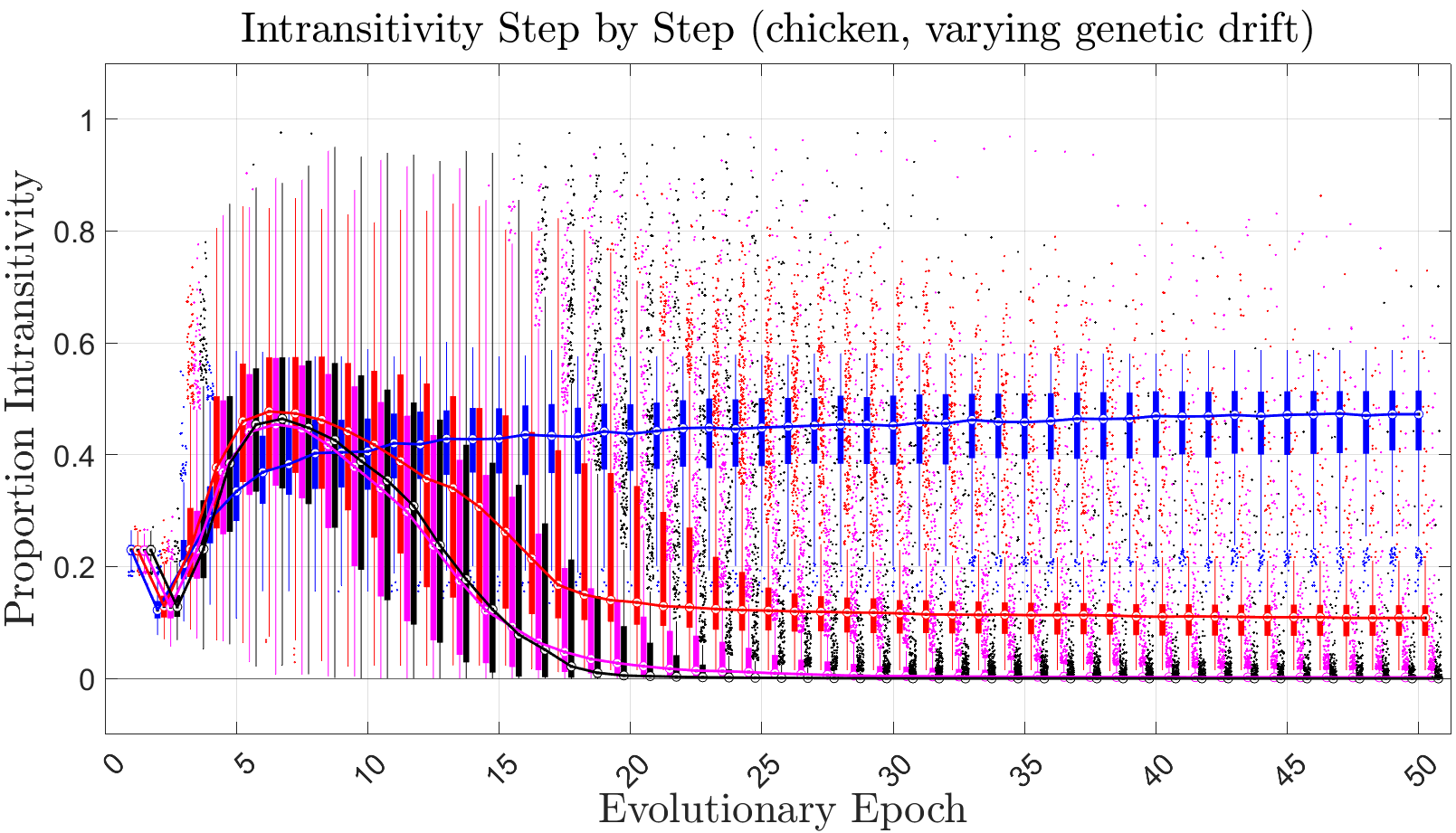}
    \caption{Step-by-step intransitivity for random performance functions. Colors corresponding to different genetic drifts: \textbf{(Blue):} $5 \cdot 10^{-2}$, \textbf{(Red):} $5 \cdot 10^{-3}$ (control), \textbf{(Magenta):} $5 \cdot 10^{-4}$, \textbf{(Black):} $5 \cdot 10^{-5}$.}
    \label{fig: chicken genetic drift}
\end{figure}

In contrast, the results using chicken are highly sensitive genetic drift parameter. With extremely small genetic drift, chicken eventually converges to near perfect transitivity, as shown in Figure \ref{fig: chicken genetic drift} (black and pink lines). With very large drift, intransitivity increases initially and stays at close to half proportion of intransitivity. The control group is approximately halfway between the two. Intransitivity eventually decreases over time, but does not converge to a negligible value.

Chicken exhibits more complex behavior since it maintains multiple clusters for longer, and the dramatic payout cusp generated by the extreme cost of collision produces large higher order derivatives. In other words, chicken is not as smooth as the preceding examples. There is likely a critical genetic drift value for chicken required for convergence to transitivity, as the performance function is far from linear on all but very small neighborhoods. Indeed, depending on the trial, we observe varying rates of convergence to transitivity depending on which derivatives dominate the local approximation to performance. Thus chicken requires the tighter concentration to a smaller portion of the trait space than the other bimatrix games to achieve convergence to transitivity.

\begin{figure}[t]
		\centering
		\includegraphics[scale=0.58]{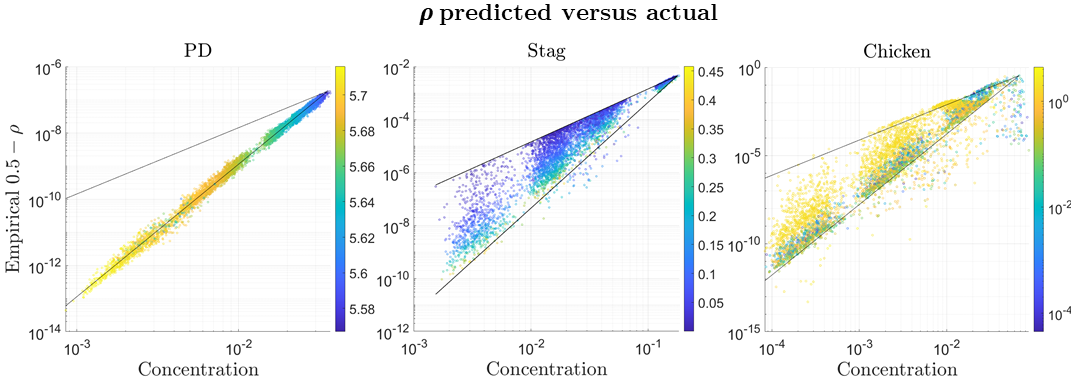}
		\caption{The empirical correlation $\rho$ versus concentration in the trait distribution. Concentration is measured by the average standard deviation across traits. We colored the points by the norm of the gradient of the performance function, $g$, for PD and stag hunt, and with $H_{xx}^2/(g  \text{ Cov})^2$ for chicken, evaluated at the final centroid. The lines mark quartic ($\mathcal{O}(\kappa^4)$) and quadratic convergence ($\mathcal{O}(\kappa^2)$) from the maximum empirical $\rho$. Data is generated using repeated trails with varying genetic drift to produce clusters with varying concentration.}
		\label{fig: stag hunt concentration vs predicted}
\end{figure}

Figure \ref{fig: stag hunt concentration vs predicted} investigates the rate of convergence of the different bimatrix games to transitivity by computing empirical estimates for the correlation $\rho$ at the end of evolution and plotting against the covariance in the final distribution. The experiment was repeated for varying levels of genetic drift to ensure that the sampled distributions span two orders of magnitude in their final covariance. The scatter plots are colored to illustrate the norms of relevant derivatives.

We compare the resulting scatter plots to the situations outlined in Figure \ref{fig:Convergence rate flowchart}. In all cases the leading order terms in the numerator are third order derivatives because the trait space is one dimensional so $H_{xy} = 0$ (see Lemma 1). We expect either quadratic or quartic convergence depending on the norm of the gradient at the centroid. When the gradent is suffcently large, convergence occurs quartically. Otherwise convergence is quadratic. We drew two lines on the figure, one indicating quadratic convergence towards $\rho = 0.5$, and one indicating quartic convergence. 

For prisoner's dilemma, we clearly observe that all of the points lie on the lower line, corresponding to quartic convergence from the maximum point. All points have nonzero gradient, so should converge at rate $\mathcal{O}(\kappa^4)$ (see Figure \ref{fig:Convergence rate flowchart}). When the gradient is nonzero the denominator is dominated by first order terms and the numerator is dominated by third order terms. Squaring the ratio leads to a quartic function in $\rho$. For stag hunt, convergence rates vary between a quadratic upper bound, and quartic lower bound. When the trait distribution converges to a location where the gradient is small (the dark blue points), convergence is quadratic since the denominator is dominated by second order terms. As the gradient grows (yellow points), the process approaches the quartic lower bound. For chicken, we color by the ratio $\frac{H_{xx}^2}{(g^2\text{Cov}[X]^2)}$ instead of by the norm of the gradient alone. We adopted this ratio since the cusp generates large higher order derivatives. Thus, the Hessian may be very large relative to the gradient, even if the gradient itself is not small. Now two clusters clearly form. When the ratio is large or near one the second order terms may dominate the denominator, so convergence may be either quadratic or quartic (interpolate between the blue and gold scenarios in the second row of Figure \ref{fig:Convergence rate flowchart}). If the ratio is small then the gradient dominates, so convergence is quartic (note the second cluster that tracks the quartic lower bound). 

For chicken, and to a lesser extent stag, some points fall below the quartic line. These points corresponded to trials whose final trait distribution contained multiple clusters of competitors. Our convergence test was not designed to treat multimodal distributions.

\subsection{Random Performance Functions} \label{sec: random performance functions}

\begin{figure}[t] 
		\centering
		\includegraphics[scale=0.6]{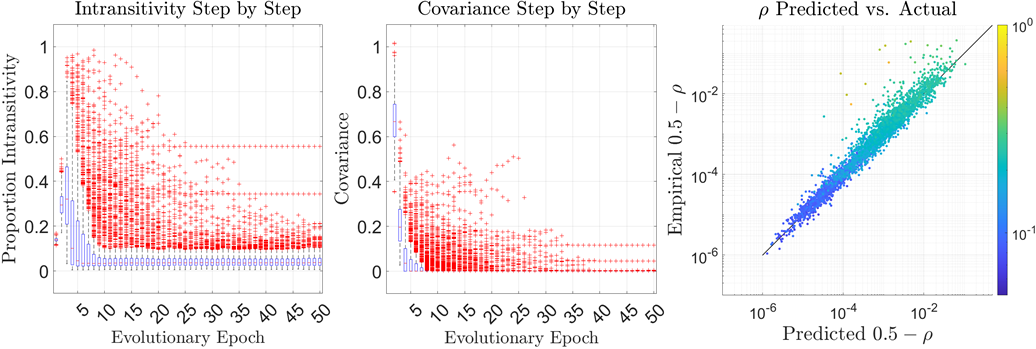}
		\caption{Trait concentration and transitivity for random performance functions under control parameters. \textbf{(Left)}: The proportion intransitivity by evolutionary epoch. \textbf{(Center):} The covariance by evolutionary epoch. \textbf{(Right):} The empirical value of 0.5 - $\rho$ compared to its prediction based on the sizes of the transitive and cyclic components. The points are colored by final covariance.}
		\label{fig: control parameters step by step}
\end{figure}

Next we extend our test to an ensemble of random performance functions chosen to illustrate the generality of our theory.

Under control parameters, intransitivity quickly vanishes. Figure \ref{fig: control parameters step by step} shows quick convergence to transitivity. Figure \ref{fig: control parameters step by step} also shows how the covariance and the number of clusters in the trait distribution evolve over time. The covariance converges towards 0 quickly, indicating rapid concentration. We also observed a quick convergence towards 1 in the number of clusters. So, as in the bimatrix examples, we observe concentration towards a delta distribution and perfect transitivity. Further, we used the convergence test outlined in Section \ref{sec: numerics} to check the predicted rates of convergence. In general, neither the gradient nor Hessian terms are 0, so we expect convergence at rate $\mathcal{O}_{=}(\kappa^2)$. We compare that prediction to our simulated results on the right of Figure \ref{fig: control parameters step by step}, and observe clear agreement across multiple orders of magnitude.

Next, we considered a series of single parameter variation tests relative to the control. Several parameters were significant, namely, the trigonometric and linear amplitudes, the dimensions of the trait space, and the amount of genetic drift. Table \ref{table: random performance functions parameters} documents the parameter variations tested.

\begin{table}[h!]
\begin{centering}
\rowcolors{2}{white!80!}{gray!10!}
\begin{tabular}{ |p{4cm}| p{4cm}|p{4cm}|  }
 \hline
 \multicolumn{3}{|c|}{List of variables of consideration} \\
 \hline
 Parameter& Control Value & Perturbations \\
 \hline
Number of competitors   & 250    & 50, 100\\
 Number of traits&   4  & 2, 8, 16, 32\\
 Number of trig. modes &2 & 4, 6\\
 Trig. mode amplitude    &1 & 0.5\\
 Linear mode amplitude&   1  & 0, 0.5\\
 Genetic drift& $5 \cdot 10^{-3}$  & $1 \cdot 10^{-3}$, $1 \cdot 10^{-2}$, $5 \cdot 10^{-2}$\\
 Games per competitor& 100  & 1, 5, 10\\
 \hline
\end{tabular}

\end{centering}
\vspace{2mm}
 \caption{A list of parameters considered in the random performance functions example.}
 \label{table: random performance functions parameters}
\end{table}

First, we altered the proportional amplitudes of the linear and trigonometric components. The trigonometric components are the primary source of roughness, and thus intransitivity, in our performance function. Therefore, decreasing their relative amplitude, should produce more transitive networks. The results of varying the amplitudes were shown in Figure \ref{fig: linear and trig amplitude intransitivity} in the Introduction (see Section \ref{sec: motivation}). There we noted the sharp decline in transitivity over time for three different parameter values to demonstrate the general concentration mechanism. We are now equipped to explain dependence on the trigonometric and linear amplitudes. Recall the difference between the red plot (control parameters) and the black plot (half trigonometric amplitude) in Figure \ref{fig: linear and trig amplitude intransitivity} (in Section \ref{sec: motivation}). In that, the black line was considerably more transitive initially because performance was more linear, however by the end of evolution both converge to near perfect transitivity.

To further explore this effect, we decreased the linear amplitude to zero producing an entirely trigonometric performance function. When the performance function is entirely trigonometric, initial competition is close to perfectly intransitive because competitors' advantage over their opponents is periodic, so most competitors are neither expected to win nor lose  against a randomly drawn opponent. Recall the blue plot in Figure \ref{fig: linear and trig amplitude intransitivity}. Despite starting out near perfect intransitivity, at the end of evolution we approach perfectly transitive competition. That sharp decline in intransitivity clearly shows that, even when we modify our game to promote extreme initial intransitivity, evolution can still promotes transitive competition via concentration.

We also increased the number of trigonometric modes in the performance function to make it rougher, and therefore, more intransitive. Increasing the number of modes slightly increased in intransitivity, but, at the end of evolution intransitivity still converged towards zero. Under a greater number of trigonometric modes, evolution took slightly slower for convergence, although not nearly as slow as for the bimatrix games. As before, even when given a performance function engineered to be more intransitive, evolution promoted near-perfect transitivity.

Next we varied the dimension of the trait space from 2 to 32 traits and compared to the control setting with 8 traits. Higher-dimensional trait spaces should admit more intransitivity because there are more opportunities for distinct traits to interact. Using 2 traits instead of 8 increased the initial transitivity, while using 32 traits, significantly decreased the initial transitivity. The more traits, the faster we observed convergence during evolution (both in the average intransitivity across trials and in the variation between trials). Broadly speaking, the fewer traits, the more variability we observed in the relative intransitivity, particularly at the end of evolution. Under all cases we observed convergence towards transitivity.

\begin{figure}[h!]
    \centering
    \includegraphics[scale=0.27]{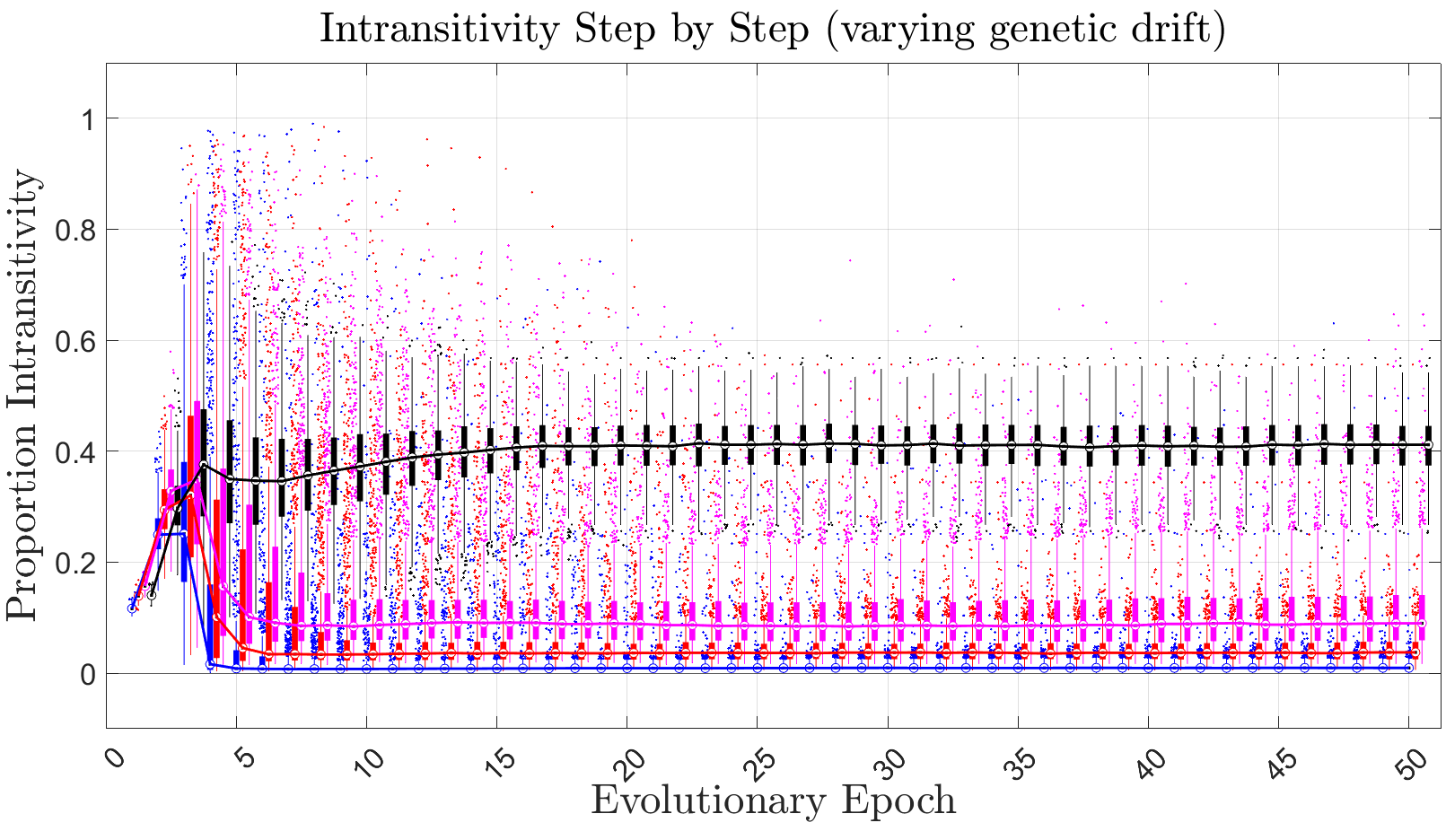}
    \caption{Step-by-step intransitivity for random performance functions. Colors corresponding to different genetic drifts: \textbf{(Blue):} $1 \cdot 10^{-3}$, \textbf{(Red):} $5 \cdot 10^{-3}$ (control), \textbf{(Magenta):} $1 \cdot 10^{-2}$, \textbf{(Black):} $5 \cdot 10^{-2}$.}
    \label{fig:genetic drift intransitivity step-by-step}
\end{figure}

In contrast, genetic drift, does affect whether the system converged to transitivity. Figure \ref{fig:genetic drift intransitivity step-by-step}, shows the results. Increasing genetic drift to 0.05 from 0.005, prevented convergence to transitivity. The resulting trait covariance converges to a small, although non-negligible, value which is an order of magnitude larger than the covariance in the control case. As a result, there is too much variation in the final trait distribution to observe convergence to transitivity. When genetic drift is lower, at 0.01, we start to see the evolutionary pattern take effect; intransitivity decreased over the course of evolution. 

All in all, we find that, while changing the parameters can change how quickly our competitive network becomes nearly perfectly transitive, the only parameter range that prevented convergence to transitivity was high genetic drift. In that case, we would not expect convergence--since evolution will not promote sufficiently concentrated trait distributions.`

%\subsection{Colonel Blotto}

%%%%%%%%%%%%%%%%%%%%%%%%%%%%%%%%%%%%%%%%%%%%%%%%%
\section{Discussion}

Our theory shows that cyclicity vanishes as traits concentrate, provided the relation between traits and performance is smooth. Our numerics show that concentration can occur in simple evolutionary models. 

Although derived without reference to a specific domain, these results have interesting consequences for specific domains. In a pairwise choice context, where performance is replaced with preference, our results demonstrate that it is, in fact, easier to compare apples to apples, than apples to oranges. Similar objects admit well ordered comparison. Diverse objects may not. In ecology, competitive cycles can maintain biodiversity by preventing competitive exclusion \cite{Reichenbach_a,Reichenbach_b,Reichenbach_c,Reichenbach_d,Soliveres,soliveres2018everything,Laird}. Here we advance the reverse hypothesis. Cycles require diversity.

Our theory established convergence in expectation to perfect transitivity given concentration. When perfectly transitive, i.e.~governed by ratings, a system is also transitive, i.e.~consistent with a rank order from best to worst. When near to perfectly transitive, most systems are also transitive \cite{strang2022network}, so it is reasonable to suspect that convergence in expectation to perfect transitivity implies convergence in probability to transitivity. Then, concentration about almost any attribute vector will almost surely lead to transitivity. We conjecture that this is true, under the assumptions that the gradient is nonzero almost everywhere, and the probability that $|f(X,Y)| > \alpha(\sigma)$ for $\alpha(\sigma) = \mathcal{O}_{\leq}(\sigma^{3/2})$ converges to one as $\sigma$ goes to zero.

The concentration mechanism requires trait spaces that are continuous enough to allow concentration in some suitable form, and performance functions that are smooth. Not all games satisfy these criteria. For example, the payout function in deterministic combinatorial games is a piecewise constant function over the set of deterministic policies. Nevertheless, exact performance function in such games are almost universally replaced with smoother evaluation functions during play, and pure policies may be replaced with stochastic decision processes, or, are themselves chosen based on a stochastic learning process (c.f.~\cite{browne2012survey}). When agent behavior is random, the distribution of possible behaviors determines expected outcomes. Then performance is a quadratic function of the underlying distribution of sampled policies. Indeed, as for any normal form game, replacing pure strategies with mixed strategies produces a game with expected payouts that are quadratic functions of the probabilities of playing each strategy. Quadratic functions are smooth, and, as illustrated in Section \ref{sec: results}, admit exact formulas relating trait covariance and intransitivity. Alternatively, games of chance, such as backgammon, naturally average similar policies over possible realizations, producing smooth performance functions (c.f.~\cite{tesauro2002programming,tesauro1995temporal}). Similar arguments ought to extend to other domains, where high level traits determine distributions of behavior and where innate randomness smooths outcomes.

Future studies could generalize our evolutionary process and concentration method. Survival probability could be continuous in fitness rather than generated via a performance cutoff. Mutation could be replaced with recombination or sexual reproduction. The overall process could also be replaced with models pertinent to a chosen domain of interest. Options include various extensions of adaptive dynamics, the replicator dynamic, ensemble methods borrowed from reinforcement learning, or population genetics models such as the Moran process. Concentration rates could be established with respect to time for specific dynamics by adapting results from \cite{Cressman_c}, or with respect to system parameters like mutation rates or noise levels using a quasi-potential approach as in \cite{foster1990stochastic,nolting2016balls,strang2020applications}.

Our findings should be tested in real-world systems. Examples from pairwise comparison studies with objects that are increasingly similar and from social hierarchies among closely related individuals offer potential test cases. In the former case, the theory may fail since individual preference among exceedingly similar objects are nearly random \cite{fudenberg1998theory}. Thus, the theory would likely only apply for preference tasks where linear models are accurate over neighborhoods that include clearly distinguishable objects. Other applications of this evolutionary framework include one-on-one competition in sports like tennis or chess, perhaps with chess engines with different styles of play. In-silico experiments allow direct parameter tuning, thus offer a well defined trait space, and an easily testable sandbox. Future work could investigate whether the concentration mechanism proposed here explains the spinning top structure observed during training in multiplayer games \cite{czarnecki2020real}.

\section{Appendices}

\subsection{Supplemental Calculations} \label{app: supplemental proofs}

\textbf{Proof of Equation \ref{eqn: quadratic approximation to performance}: } Equation \ref{eqn: quadratic approximation to performance} states that, up to second order:
$$
    f(x,y) \simeq r(x|z) - r(y|z) + (x - z)^{\intercal} H_{xy}(z,z) (y - z) + \mathcal{O}((x - z, y - z)^3). 
$$
where:
$$
    r(x|z) = \nabla_x f(z,z)^{\intercal} (x - z) + \frac{1}{2}(x - z)^{\intercal} H_{xx}(z,z) (x - z). 
$$

The proof follows from the alternating structure of the derivatives. Accounting for the alternating derivatives, equation \ref{eqn: quadratic not simplified} becomes:
$$
\begin{aligned}
    f(x,y) \simeq & \nabla_{x} f(z,z)^{\intercal} (x - y) + \frac{1}{2} (x - z)^{\intercal} H_{xx}(x,y) (x - z) - \frac{1}{2} (y-z)^{\intercal} H_{yy} (y - z) + \hdots \\
    & + \frac{1}{2} \left[ (x - z)^{\intercal} H_{xy}(z,z) (y - z) - (y - z) H_{xy}(z,z) (x - z) \right] + \mathcal{O}((x - z, y - z)^3).
\end{aligned}
$$
where the simplifications follow from replacing $\nabla_y f(z,z)$ with $- \nabla_x f(z,z)$, $H_{yy}(z,z)$ with $-H_{xx}(z,z)$, and $H_{yx}(z,z)$ with $- H_{xy}(z,z)$. Then, since $H_{xy}$ is skew symmetric:
$$
    (y - z)^{\intercal} H_{xy} (x - z) = \left((y - z)^{\intercal} H_{xy} (x - z) \right)^{\intercal} = (x - z)^{\intercal} H^{\intercal}_{xy} (y - z) = -(x - z)^{\intercal} H_{xy}(z,z) (y - z).
$$
So:
$$
\begin{aligned}
    f(x,y) \simeq & \nabla_{x} f(z,z)^{\intercal} (x - y) + \frac{1}{2} (x - z)^{\intercal} H_{xx}(x,y) (x - z) - \frac{1}{2} (y-z)^{\intercal} H_{yy} (y - z) + \hdots \\
    &  +  (x - z)^{\intercal} H_{xy}(z,z) (y - z) + \mathcal{O}((x - z, y - z)^3).
\end{aligned}
$$

Then:
$$
    f(x,y) \simeq r(x|z) - r(y|z) + (x - z)^{\intercal} H_{xy}(z,z) (y - z) + \mathcal{O}((x - z, y - z)^3). 
$$
where:
$$
    r(x|z) = \nabla_x f(z,z)^{\intercal} (x - z) + \frac{1}{2}(x - z)^{\intercal} H_{xx}(z,z) (x - z). \quad \blacksquare
$$

\vspace{6mm}

%%%%%%%%%%%%%%%%%%%%%%%%%%%%%%%%%%%%%%%%

\textbf{Proof of Equation  \ref{eqn: rho in expectations}:} Equation \ref{eqn: rho in expectations} states that the correlation coefficient is given by:
$$
    \rho = \frac{\mathbb{V}_{X}[r(X)] + 2 \mathbb{E}_{X,Y}[r(X) h(X,Y)] + \mathbb{E}_{X}\left[ \mathbb{E}_{Y}[h(X,Y)]^2 \right] }{2 \mathbb{V}_{X}[r(X)] + 4 \mathbb{E}_{X,Y}[r(X) h(X,Y)] + \mathbb{E}_{X,Y}[h(X,Y)^2]}
$$

To start we simplify the expected performance. Substituting in for $f$:
$$
\int_{\Omega} f(x,y) \pi_x(y) dy = \int_{\Omega} (r(x) - r(y) + h(x,y) ) \pi_x(y) dy  = r(x) - \int_{\Omega} r(y) \pi_x(y) dy + \int_{\Omega} h(x,y) \pi_x(y) dy.
$$

Now, by choice of the constant added to $r(x)$, the expected performance is:
$$
\int_{\Omega} f(x,y) \pi_x(y) dy = r(x) + \int_{\Omega} h(x,y) \pi_x(y) dy.
$$

To compute the uncertainty in the expected performance square the expected performance and integrate:
$$
\begin{aligned}
 &\int_{\Omega} \left( \int_{\Omega} f(x,y) \pi_x(y) dy \right)^2 \pi_x(x) dx  = \int_{\Omega} \left(r(x) +  \int_{\Omega} h(x,y) \pi_x(y) dy \right)^2 \pi_x(x) dx \\
 & \hspace{1.5 cm} = \int_{\Omega} r(x)^2 \pi_x(x) dx + 2 \int_{\Omega}\int_{\Omega} r(x) h(x,y) \pi_x(y) \pi_x(x) dy dx + 
 \int_{\Omega} \left( \int_{\Omega} h(x,y) \pi_x(y) dy \right)^2 \pi_x(x) dx.
\end{aligned}
$$

Now, going term by term, $\mathbb{E}_{X}[r(X)] = 0$ so $\int_{\Omega} r(x)^2 \pi_x(x) dx = \mathbb{V}_{x}[r(X)]$ is the variance in the rating of a randomly drawn competitor, $r(X)$. The next term is $2 \mathbb{E}_{X,Y}[r(X) h(X,Y)]$ and the last term is $\mathbb{E}_{X,Y}[h(X,Y)^2]$. Therefore:
$$
\int_{\Omega} \left( \int_{\Omega} f(x,y) \pi_x(y) dy \right)^2 \pi_x(x) dx = \mathbb{V}_{X}[r(X)] + 2 \mathbb{E}_{X,Y}[r(X) h(X,Y)] + \mathbb{E}_{X}\left[ \mathbb{E}_{Y}[h(X,Y)]^2 \right].
$$

Next consider the uncertainty in performance:
$$
\begin{aligned}
&\int_{\Omega}\int_{\Omega} f(x,y)^2 \pi_x(y) \pi_x(x) dy dx = \int_{\Omega}\int_{\Omega} (r(x) - r(y) + h(x,y))^2 \pi_x(y) \pi_x(x) dy dx \\
& \hspace{1.5 cm} = \int_{\Omega}\int_{\Omega} \left(r(x)^2 +r(y)^2 - 2 r(x) r(y) + 2 r(x) h(x,y) - 2 r(y) h(x,y) + h(x,y)^2 \right) \pi_x(y) \pi_x(x) dy dx. 
\end{aligned}
$$

The first terms are identical and both equal $\mathbb{V}_{X}[r(X)]$. The next cross term $\int_{\Omega}\int_{\Omega} r(x) r(y) \pi_x(x) \pi_x(y) dx dy = \mathbb{E}_{X,Y}[r(X) r(Y)] = 0$ since $X$ and $Y$ are independent and identically distributed, so $\mathbb{E}_{X,Y}[r(X) r(Y)] = \mathbb{E}_{X}[r(X)]^2$ and $\mathbb{E}_{X}[r(X)] = 0$.

The next pair of cross terms are also identical since $h(x,y)$ also obeys the fairness criterion. By fairness:
$$
r(x) - r(y) + h(x,y) = -\left( r(y) - r(x) + h(y,x) \right) = r(x) - r(y) - h(y,x)
$$
so $h(x,y) = -h(y,x)$. Then $-\mathbb{E}_{X,Y}[r(Y) h(X,Y)] = \mathbb{E}_{X,Y}[r(Y) h(Y,X)] = \mathbb{E}_{X,Y}[r(X) h(X,Y)]$.

Therefore, the uncertainty in performance is:
$$
\int_{\Omega}\int_{\Omega} f(x,y)^2 \pi_x(y) \pi_x(x) dy dx = 2 \mathbb{V}_{X}[r(X)] + 4 \mathbb{E}_{X,Y}[r(X) h(X,Y)] + \mathbb{E}_{X,Y}[h(X,Y)^2].
$$

Thus:
$$
    \rho = \frac{\mathbb{V}_{X}[r(X)] + 2 \mathbb{E}_{X,Y}[r(X) h(X,Y)] + \mathbb{E}_{X}\left[ \mathbb{E}_{Y}[h(X,Y)]^2 \right] }{2 \mathbb{V}_{X}[r(X)] + 4 \mathbb{E}_{X,Y}[r(X) h(X,Y)] + \mathbb{E}_{X,Y}[h(X,Y)^2]}. \quad \blacksquare
$$

\vspace{4mm}
%%%%%%%%%%%%%%%%%%%%%%%%%

\textbf{Proof of Equations \ref{eqn: variance in ratings} and \ref{eqn: expected h squared}:} Equations \ref{eqn: variance in ratings} and \ref{eqn: expected h squared} evaluate the expectation required to compute $\epsilon$ explicitly provided $f(x,y)$ is quadratic. Those calculations are performed below.

Since $r$ is mean zero its variance is:
$$
\begin{aligned}
    \mathbb{V}_{X}[r(X)] = \mathbb{E}_{X}[r(X)^2] = \mathbb{E}_X\left [\left(\nabla_x f(z,z)^{\intercal} (X - z) + \frac{1}{2} (X - z)^{\intercal} H_{xx} (X - z) \right)^2 \right] - \bar{r}^2.
\end{aligned}
$$
where $\bar{r} = \mathbb{E}_{X}[r(X)]$. 

To evaluate the expectation, expand the square:
$$
\begin{aligned}
& \mathbb{E}_X\left [\left(\nabla_x f(z,z)^{\intercal} (X - z) + \frac{1}{2} (X - z)^{\intercal} H_{xx} (X - z) \right)^2 \right] = \\ & \hspace{6mm} = \mathbb{E}_{X}\left[ \left(\nabla_x f(z,z)^{\intercal} (X - z) \right)^2  \right] + \mathbb{E}_{X}\left[ \left(\nabla_x f(z,z)^{\intercal} (X - z) \right) (X - z)^{\intercal} H_{xx} (X - z) \right] + \hdots \\ & \hspace{6mm} + \frac{1}{4} \mathbb{E}_{X}\left[ \left( (X - z)^{\intercal} H_{xx} (X - z) \right)^2 \right]
\end{aligned}
$$

The first term is the easiest to evaluate:
$$
    \mathbb{E}_{X}\left[ \left(\nabla_x f(z,z)^{\intercal} (X - z) \right)^2  \right] = \sum_{i,j} \partial_{x_i} f \partial_{x_j} f \mathbb{E}[(X - z)_i (X - z)_j] = \nabla_{x} f(z,z)^{\intercal} \text{Cov}[X] \nabla_x f = \langle \nabla_x f (\nabla_x f)^{\intercal}, \text{Cov}[X] \rangle.
$$

To help simplify our notation we will adopt Einstein summation notation. Then:
$$
\mathbb{E}_{X}\left[ \left(\nabla_x f(z,z)^{\intercal} (X - z) \right)^2  \right] = \nabla_x f^{\alpha} \nabla_x f^{\beta} \text{Cov}[X]_{\alpha \beta}
$$
where $\alpha, \beta$ stand for a range of indices associated with the given array, and we evaluate the sum over any pair of indices that appear in a super and subscript. In this notation:
$$
\bar{r} = \frac{1}{2} H_{xx}^{\alpha \beta} \text{Cov}[X]_{\alpha \beta}.
$$
so:
$$
\bar{r}^2 = \frac{1}{4} H_{xx}^{\alpha \beta} H_{xx}^{\mu \nu} \text{Cov}[X]_{\alpha \beta} \text{Cov}[X]_{\mu \nu}.
$$

To finish computing the variance in the local rating function we expand the remaining terms:
$$
\begin{aligned}
\mathbb{E}_{X} \left[ \left(\nabla_x f(z,z)^{\intercal} (X - z) \right) (X - z)^{\intercal} H_{xx} (X - z) \right] &  = \sum_{i,j,k} (\partial_{x_i} f(z,z)) (\partial_{x_j} \partial_{x_k} f(z,z)) \mathbb{E}_{X}[(X - z)_i (X - z)_j (X - z)_k] \\ & = \nabla f(z,z)^{\alpha} H_{xx}^{\mu \nu} \text{Th}[X]_{\alpha \mu \nu}
\end{aligned}
$$
where $\text{Th}[X]$ is the tensor of third order central moments of $X$. 

The last term to expand is:
$$
\begin{aligned}
\mathbb{E}_{X}\left[ \left( (X - z)^{\intercal} H_{xx} (X - z) \right)^2 \right] & = \sum_{i,j,k,l} (\partial_{x_i} \partial_{x_j} f) (\partial_{x_k} \partial_{x_l} f) \mathbb{E}_{X}[(X - z)_i (X - z)_j (X - z)_k (X - z)_l] \\ 
& = H_{xx}^{\alpha \beta} H_{xx}^{\mu \nu} \text{Fo}[X]_{\alpha \beta \mu \nu}
\end{aligned}
$$
where $\text{Fo}[X]$ is the tensor of central fourth moments.

Therefore:
$$
    \mathbb{V}_X[r(X)] = \nabla_x f(z,z)^{\alpha} \nabla_x f(z,z)^{\beta} \text{Cov}[X]_{\alpha \beta} + \nabla f(z,z)^{\alpha} H_{xx}^{\mu \nu} \text{Th}[X]_{\alpha \mu \nu} + \frac{1}{4} H_{xx}^{\alpha \beta} H_{xx}^{\mu \nu} \left( \text{Fo}[X]_{\alpha \beta \mu \nu} - \text{Cov}[X]_{\alpha \beta} \text{Cov}[X]_{\mu \nu} \right) .
$$

All that is left then is to expand the numerator of $\epsilon$:
$$
\begin{aligned}
\mathbb{E}_{X,Y}[h(X,Y)^2] & = \sum_{i,j,k,l} (\partial_{x_i} \partial_{x_j} f) (\partial_{x_k} \partial_{x_l} f) \mathbb{E}_{X,Y}[(X - z)_i (Y - z)_j (X - z)_k (Y - z)_l] \\ & = \sum_{i,j,k,l} (\partial_{x_i} \partial_{y_j} f) (\partial_{x_k} \partial_{y_l} f) \mathbb{E}_X[(X - z)_i (X - z)_k] \mathbb{E}_Y[ (Y - z)_j  (Y - z)_l] \\
& = H_{xy}^{\alpha \beta} H_{xy}^{\mu \nu} \text{Cov}[X]_{\alpha \mu} \text{Cov}[X]_{\beta \nu}
\end{aligned}
$$
where the second to last equality follows from the independence of $X$ ad $Y$, and the last equality follows from the fact that they are identically distributed. Therefore\footnote{We will see this sort of tensor product again, so it will be helpful to be able to write the product using more straightforward linear algebra. By rearranging the sums involved in a product $A^{\alpha \beta} A^{\mu \nu} B_{\alpha \mu} B_{\beta \nu}$ one can show that the product equals $\text{trace}(A B A^{\intercal} B)$. This is easy to check:
$$
\begin{aligned}
\text{trace}(A B A^{\intercal} B^{\intercal}) & = \sum_{i} (A B A^{\intercal} B^{\intercal})_{i,i} = \sum_{i} \sum_{l} (A B)_{i,l} (B^{\intercal} A^{\intercal})^{\intercal}_{l,i} & \\
& = \sum_{i,k} \left(\sum_{j} A_{i,j} B_{j,l} \right) \left(\sum_{k} A^{\intercal}_{l,k} B^{\intercal}_{k,i} \right)  = \sum_{i,j,k,l} A_{i,j} B_{j,l} A_{k,l} B_{i,k} \\
& = \sum_{i,j,k,l} A_{i,j} A_{k,l} B_{i,k} B_{j,l}.
\end{aligned}
$$
It follows that:
\begin{equation}
    A^{\alpha \beta} A^{\mu \nu} B_{\alpha \mu} B_{\beta \nu} = \text{trace}(A B A^{\intercal} B^{\intercal}) = \langle A B, B A \rangle. 
\end{equation}
This equation makes practical implementation of the tensor product easier. It also makes it easier to analyze. 
}:
$$
    \mathbb{E}_{X,Y}[h(X,Y)^2] = H_{xy}^{\alpha \beta} H_{xy}^{\mu \nu} \text{Cov}[X]_{\alpha \mu} \text{Cov}[X]_{\beta \nu} \quad \blacksquare
$$

\vspace{4mm}
%%%%%%%%%%%%%%%%%%%%%%%%%%%%%%%%%%%%%%%%
\textbf{Proof of Equation \ref{eqn: epsilon normal}:}

Suppose $X \sim \mathcal{N}(z,\Sigma)$. Then:
\begin{equation}
\text{Fo}[X]_{i,j,k,l} = \sigma_{ij} \sigma_{kl} + \sigma_{ik} \sigma_{jl} + \sigma_{il} \sigma_{jk}. 
\end{equation}

Then:
$$
\begin{aligned}
H_{xx}^{\alpha \beta} H_{xx}^{\mu \nu} \left(\text{Fo}[X]_{\alpha \beta \mu \nu} - \text{Cov}[X]_{\alpha \beta} \text{Cov}[X]_{\mu \nu} \right) & = H_{xx}^{\alpha \beta} H_{xx}^{\mu \nu} \left(\Sigma_{\alpha \beta} \Sigma_{\mu \nu} + \Sigma_{\alpha \mu} \Sigma_{\beta \nu} + \Sigma_{\alpha \nu} \Sigma_{\beta \mu} - \Sigma_{\alpha \beta} \Sigma_{\mu \nu} \right) \\
& = H_{xx}^{\alpha \beta} H_{xx}^{\mu \nu} \left( \Sigma_{\alpha \mu} \Sigma_{\beta \nu} + \Sigma_{\alpha \nu} \Sigma_{\beta \mu}  \right)
\end{aligned}
$$

Now, since $H_{xx}$ is symmetric $H_{xx}^{\mu \nu} = H_{xx}^{\nu \mu}$:
$$
H_{xx}^{\alpha \beta} H_{xx}^{\mu \nu} \Sigma_{\alpha \mu} \Sigma_{\beta \nu} =  H_{xx}^{\alpha \beta} H_{xx}^{\nu \mu} \Sigma_{\alpha \nu} \Sigma_{\beta \mu}  = H_{xx}^{\alpha \beta} H_{xx}^{\mu \nu} \Sigma_{\alpha \mu} \Sigma_{\beta \nu}
$$
where the last equality follows from renaming the indices $\nu = \mu$ and $\mu = \nu$. It follows that:
$$
\frac{1}{4} H_{xx}^{\alpha \beta} H_{xx}^{\mu \nu} \left(\text{Fo}[X]_{\alpha \beta \mu \nu} - \text{Cov}[X]_{\alpha \beta} \text{Cov}[X]_{\mu \nu} \right)  =  \frac{1}{2} H_{xx}^{\alpha \beta} H_{xx}^{\mu \nu}  \Sigma_{\alpha \mu} \Sigma_{\beta \nu}
$$

Therefore:
$$
\epsilon = \frac{H_{xy}^{\alpha \beta} H_{xy}^{\mu \nu} \Sigma_{\alpha \mu} \Sigma_{\beta \nu}}{2 \nabla_x f(z,z)^{\alpha} \nabla_x f(z,z)^{\beta} \Sigma_{\alpha \beta} +  H_{xx}^{\alpha \beta} H_{xx}^{\mu \nu} \Sigma_{\alpha \mu} \Sigma_{\beta \nu} }.
$$

\vspace{4mm}
%%%%%%%%%%%%%%%%%%%%%%%%%%%%%%%%%%%%%%%%
\textbf{Proof of Equation \ref{eqn: epsilon Boltsmann}} Equation \ref{eqn: epsilon Boltsmann} describes the value of $\epsilon$ when performance is quadratic, and the traits are normally distributed with covariance $\Sigma \propto H_{xx}^{-1}$ about some centroid where the gradient in performance vanishes. 

Since the covariance $\Sigma$ appears in the numerator and denominator of $\epsilon$ the constant of proportionality relating $\Sigma$ and $H_{xx}^{-1}$ does not influence $\epsilon$. Therefore:
$$
\epsilon = \frac{H_{xy}^{\alpha \beta} H_{xy}^{\mu \nu} {H_{xx}^{-1}}_{\alpha \mu} {H_{xx}^{-1}}_{\beta \nu}}{H_{xx}^{\alpha \beta} H_{xx}^{\mu \nu} {H_{xx}^{-1}}_{\alpha \mu} {H_{xx}^{-1}}_{\beta \nu}}
$$

In this case the tensor product in the denominator simplifies since:
$$
\begin{aligned}
{H_{xx}^{-1}}_{\alpha \mu} {H_{xx}^{-1}}_{\beta \nu} & = \sum_{i,j} \sum_{k,l} {H_{xx}}_{i,j} {H_{xx}}_{k,l} {H_{xx}^{-1}}_{i,k} {H_{xx}^{-1}}_{j,l} = \sum_{i,j,k} {H_{xx}}_{i,j} {H_{xx}^{-1}}_{i,k} \sum_{l} {H_{xx}}_{k,l} {H_{xx}^{-1}}_{j,l} \\
& = \sum_{i,j,k} {H_{xx}}_{i,j} {H_{xx}^{-1}}_{i,k} [H_{xx} H_{xx}^{-\intercal}]_{k,j} =  \sum_{i,j,k} {H_{xx}}_{i,j} {H_{xx}^{-1}}_{i,k} [H_{xx} H_{xx}^{-1}]_{k,j} = \sum_{i,j,k} {H_{xx}}_{i,j} {H_{xx}^{-1}}_{i,k} \delta_{k,j} \\ 
& = \sum_{i,j} {H_{xx}}_{i,j} {H^{-1}_{xx}}_{i,j} = \text{trace}[H_{xx} H_{xx}^{-\intercal}] = \text{trace}[H_{xx} H_{xx}^{-1}] = T 
\end{aligned}
$$
where $T$ is the dimension of the trait space. 

Then:
$$
    \epsilon = \frac{1}{T} H_{xy}^{\alpha \beta} H_{xy}^{\mu \nu} {H_{xx}^{-1}}_{\alpha \mu} {H_{xx}^{-1}}_{\beta \nu} = \frac{1}{T} \langle H_{xy} H_{xx}^{-1}, H_{xx}^{-1} H_{xy} \rangle. \quad \blacksquare
$$

%%%%%%%%%%%%%%%%%%%%%%%%%%%%%%%%%%%%%%%%%%
\vspace{6 mm}

\textbf{Proof of Lemma 2: } If performance is quadratic then:
\begin{equation}
    h(x,y|z) = (x - z)^{\intercal} H_{xy}(z,z) (y - z).
\end{equation}
where $H_{xy}(z,z)$ is skew symmetric. If performance is quadratic then the local quadratic approximation about any $z$ is the same as the performance function, so is independent of $z$. Thus we are free to choose $z$ arbitrarily. Let $z = \mathbb{E}_X[X]$. Then:
\begin{equation}
    \mathbb{E}_{Y}[h(X,Y|z)] = (X - z) H_{x,y}(z,z) \mathbb{E}_{Y}[Y - z] = 0. 
\end{equation}

Thus the term in the numerator of Equation \ref{eqn: epsilon} which was set to zero to give a lower bound on $\rho$ is zero for quadratic performance functions, so the bound on $\rho$ in terms of $\epsilon$ is tight for all quadratic performance functions. 

Then, to show that $\epsilon$ simplifies note that:
$$
\mathbb{E}_{X,Y}[r(X|z) h(X,Y|z)] = \mathbb{E}_{X}[r(X|z) \mathbb{E}_{Y}[h(X,Y|z)]] = 0.
$$
Thus, the denominator in $\rho$ reduces to twice the variance in the local rating function.
$\blacksquare$

\vspace{6 mm}

%%%%%%%%%%%%%%%%%%%%%%%%%%%%
\textbf{Proof of Lemma 3: } If performance is quadratic then:
$$
\epsilon = \frac{1}{2}\ltfrac{H_{xy}^{\alpha \beta} H_{xy}^{\mu \nu} \text{Cov}[X]_{\alpha \mu} \text{Cov}[X]_{\beta \nu}}{\nabla_x f(z,z)^{\alpha} \nabla_x f(z,z)^{\beta} \text{Cov}[X]_{\alpha \beta} + \nabla f(z,z)^{\alpha} H_{xx}^{\mu \nu} \text{Th}[X]_{\alpha \mu \nu} + \frac{1}{4} H_{xx}^{\alpha \beta} H_{xx}^{\mu \nu} \left(\text{Fo}[X]_{\alpha \beta \mu \nu} - \text{Cov}[X]_{\alpha \beta} \text{Cov}[X]_{\mu \nu}  \right) }.
$$

Consider the denominator first. The leading term in the denominator is:
$$
\nabla_x f(z,z)^{\alpha} \nabla_x f(z,z)^{\beta} \text{Cov}[X]_{\alpha \beta} = \nabla_x f(z,z)^{\intercal}  \text{Cov}[X] \nabla_x f(z,z) \in [\sigma_{\min}(\text{Cov}[X]), \sigma_{\max}(\text{Cov}[X])] \| \nabla_x f(z,z)^{\beta} \|^2.
$$

Then, under the assumption that $\text{Cov}[X]$ is $\mathcal{O}_{=}(\kappa^2)$, so are its singular values, so the leading term in the denominator is $\mathcal{O}_{=}(\kappa^2
)$. The remaining terms all involve higher order moments which are, by assumption, $\mathcal{O}_{<}(\kappa^2)$, so are dominated by the leading term in the limit and vanish. Then the numerator is $\mathcal{O}_{=}(\kappa^2)$. 

Let:
\begin{equation}
    \Sigma = \lim_{\kappa \rightarrow 0} \frac{1}{\kappa^2} \text{Cov}[X].
\end{equation}

Then:
$$
\begin{aligned}
\lim_{\kappa \rightarrow 0} \frac{1}{\kappa^2} \epsilon & = \lim_{\kappa \rightarrow 0} \frac{1}{2} \frac{\kappa^{-4}}{\kappa^{-2}} \frac{ H_{xy}^{\alpha \beta} H_{xy}^{\mu \nu} \text{Cov}[X]_{\alpha \mu} \text{Cov}[X]_{\beta \nu}}{ \left( \nabla_x f(z,z)^{\alpha} \nabla_x f(z,z)^{\beta} \text{Cov}[X]_{\alpha \beta} + \mathcal{O}_{<}(\kappa^2) \right)}
 = \frac{1}{2} \frac{ H_{xy}^{\alpha \beta} H_{xy}^{\mu \nu} \Sigma_{\alpha \mu} \Sigma_{\beta \nu}}{  \nabla_x f(z,z)^{\alpha} \nabla_x f(z,z)^{\beta} \Sigma_{\alpha \beta} }.
\end{aligned}
$$

The limit is nonzero provided the numerator is nonzero. The numerator equals $\|\tilde{H}_{xy}\|_{\text{Fro}}^2$ in the coordinate system that whitens the limiting trait distribution. That is, the numerator is the Frobenius norm of the $xy$ block of the Hessian after changing coordinates to send $\Sigma$ to the identity. As long as $\Sigma$ is invertible the associated transform is a similarity transform so $\tilde{H}_{xy} = 0$ if and only if $H_{xy}$. The Frobenius norm of a matrix is zero if and only if the matrix itself is zero, therefore the numerator is nonzero provided $\Sigma$ is invertible, and $H_{xy} \neq 0$. The limiting covariance $\Sigma$ is nonsingular by the assumption that $\text{Cov}[X]$is $\mathcal{O}_{=}(\kappa^2)$ so the numerator is nonzero if $H_{xy}$ is nonzero, and zero otherwise. Thus:
$$
\epsilon = \mathcal{O}_{\leq}(\kappa^2)
$$
with equality if and only if $H_{xy} \neq 0$.

Then, to establish the convergence results for the expected sizes of the components note that:
\begin{equation}
    \frac{\mathbb{E}[||F_c||^2]}{\mathbb{E}[||F||^2]} = \frac{\epsilon}{1 + \epsilon} \frac{L}{E} = \mathcal{O}(\kappa^2). \quad \blacksquare
\end{equation}

\vspace{6 mm}

%%%%%%%%%%%%%%%%%%%%%%%%%%%%%%%%%%%%%%%%%%%%%%%

\textbf{Proof of Lemma 4: } The variance in performance is the denominator in the ratio defining $\rho$ (see equation \ref{eqn: rho integral}), which we have already computed for quadratic functions. When performance is quadratic:
$$
\begin{aligned}
\mathbb{V}_{X,Y}[f(X,Y)] & = 2 \mathbb{V}_{X}[r(X|z)] + 4 \mathbb{E}_{X,Y}[r(X|z) h(X,Y|z)] + \mathbb{E}_{X,Y}[h(X,Y|z)^2] \\
& = (2 \mathbb{V}_{X}[r(X|z)] + 4 \mathbb{E}_{X,Y}[r(X|z) h(X,Y|z)]) \left(1 + \epsilon \right)
\end{aligned}
$$

We have already shown that in the quadratic case $\mathbb{E}_{X,Y}[r(X|z) h(X,Y|z)] = 0$ and $\epsilon$ is $\mathcal{O}_{\leq}(\kappa^2)$ so:
\begin{equation}
    \mathbb{V}_{X,Y}[f(X,Y)] = 2 \mathbb{V}_{X}[r(X|z)](1 + \mathcal{O}_{\leq}(\kappa^2)).
\end{equation}

The variance in the local rating function is (see Equation \ref{eqn: variance in ratings}):
$$
\begin{aligned}
     \mathbb{V}_{X}[r(X|z)] = &  \nabla_x f(z,z)^{\alpha} \nabla_x f(z,z)^{\beta} \text{Cov}[X]_{\alpha \beta} + \nabla f(z,z)^{\alpha} H_{xx}^{\mu \nu} \text{Th}[X]_{\alpha \mu \nu} + \hdots \\ & \hdots + \frac{1}{4} H_{xx}^{\alpha \beta} H_{xx}^{\mu \nu} \left( \text{Fo}[X]_{\alpha \beta \mu \nu} - \text{Cov}[X]_{\alpha \beta} \text{Cov}[X]_{\mu \nu} \right) 
\end{aligned}
$$

Under the assumptions of the Lemma, $\text{Cov}[X]$ is $\mathcal{O}_{=}(\kappa^2)$, so the first term in the variance in ratings is $\mathcal{O}_{=}(\kappa^2)$. Under the assumptions of the Lemma the third and fourth moments all vanish at least as fast as the covariance in the concentration parameter. Thus, the variance in the ratings is $\mathbb{V}_X[r(X|z)] = \mathcal{O}_{=}(\kappa^2)$. 

So, if performance is quadratic, the trait distribution has mean $z$, has covariance $\mathcal{O}(\kappa^2)$, and third and fourth order moments that vanish at least as fast as the covariance, then:
$$
    \mathbb{V}_{X,Y}[f(X,Y)] = \mathcal{O}_{=}(\kappa^2)
$$
so:
$$
\begin{aligned}
    & \mathbb{E}[||F||^2] = \mathcal{O}_{=}(\kappa^2) \\
    & \mathbb{E}[||F_t||^2] = \mathcal{O}_{=}(\kappa^2) \\
    & \mathbb{E}[||F_c||^2] = \mathcal{O}_{\leq}(\kappa^4) \\   
\end{aligned}
$$
thereby proving the Lemma statement. $\blacksquare$

\vspace{6mm} 

%%%%%%%%%%%%%%%%%%%%%%%%%%%%%%%%%%%%%%%%%%%%%%%%%
\textbf{Proof of Lemma 6: } Suppose that, for any $z$, the Taylor expansion of $f(x,y)$ about $z$ converges at all $x,y \in \Omega \times \Omega$. Then $f(x,y)$ can be replaced by its Taylor series expansion about $z$ at any $z$ without introducing any error. Then the discrepancy between $f(x,y)$ and its quadratic approximation about $z$, $g(x,y|z)$, consists of the third and higher order terms in the series expansion of $f$. Then all of the products involving $g(X,Y|z)$ can be expanded into a power series in $(X - z), (Y - z)$. Then all of the products appearing in the terms introduced by $g$ could be expanded into a linear combination of the central moments of the trait distribution. Then, to understand the behavior of these terms we have to understand the behavior of the central moments in the concentration limit $\kappa \rightarrow 0$. 

Let $\alpha = [\alpha_1,\alpha_2,\hdots, \alpha_T]$ be a multi-index and let $|\alpha| = \sum_{j} \alpha_j$. Then let $x^{\alpha} = \prod_{j=1}^T x_j^{\alpha_j}$. Then, by assumption:
\begin{equation}
    \mathbb{E}_{X \sim \pi_x}[(X - z)^{\alpha}] = \begin{cases} 1 \text{ if } |\alpha| = 0 \\
    0 \text{ if } |\alpha| = 1 \\
    \mathcal{O}_{\leq}(\kappa^{|\alpha|}) \text{ if } |\alpha| \in [2,3,4,5]\\
    \mathcal{O}_{<}(\kappa^5) \text{ if } |\alpha| > 5
    \end{cases}
\end{equation}

The power series expansion of $f$ involves products of $x - z$ and $y - z$. In an expectation these can be expressed $\mathbb{E}_{X,Y \sim \pi_x}[(X -z)^{\alpha} (Y - z)^{\beta}]$ for some pair of multi-indices $\alpha$ and $\beta$. Since $X$ and $Y$ are sampled independent, any central moment in $X$ and $Y$ can be expressed as a product of moments, $\mathbb{E}_{X \sim \pi_x}[(X -z)^{\alpha}] \mathbb{E}_{Y\sim \pi_x}[(Y - z)^{\beta}]$. Then, since $X$ and $Y$ are sampled indentically, $\mathbb{E}_{X,Y \sim \pi_x}[(X -z)^{\alpha} (Y - z)^{\beta}]$ equals the product $\mathbb{E}_{X \sim \pi_x}[(X -z)^{\alpha}] \mathbb{E}_{X\sim \pi_x}[(X - z)^{\beta}]$. 

Next, recall that $\epsilon$ is given by:
$$
\epsilon = \frac{\mathbb{E}_{X,Y}[\left((X - z)^{\intercal} H_{xy}(z,z) (Y - z) \right)^2] + 2 \mathbb{E}[(X - z)^{\intercal} H_{xy}(z,z) (Y - z) g(X,Y|z)] + \mathbb{E}_{X,Y}[g(X,Y|z)^2]}{2 \mathbb{V}_{X}[r(X|z)] + 4 \mathbb{E}_{X,Y}[r(X|z) g(X,Y|z)]}.
$$

Then, to bound the rate of convergence of $\epsilon$ to zero we identify the lowest order moments in each of the expectations used to define $\epsilon$. We have already derived these expansions for the leading term in the numerator and denominator (see the quadratic analysis). Therefore we need to identify the lowest order moments in:
$$
\begin{aligned}
& \mathbb{E}_{X,Y}[r(X|z) g(X,Y|z)] \\
& \mathbb{E}_{X,Y}[(X - z)^{\intercal} H_{xy}(z,z) (Y - z) g(X,Y|z)] \\
& \mathbb{E}_{X,Y}[g(X,Y|z)^2]. \\
\end{aligned}
$$

First, $r(x|z)$ consists of a linear term in $x - z$, a quadratic term in $x - z$, a zeroeth order term that involves a tensor product between $H_{xx}$ and $\text{Cov}[X]$, which is $\mathcal{O}_{=}(\kappa^2)$, so has entries that are $\mathcal{O}_{\leq}(\kappa^2)$. The lowest order terms in $g$ are cubic. Thus, all terms in $g$ are $\mathcal{O}_{<}(\kappa^2)$. Thus, the lowest order moments in $r(X|z) g(X,Y|z)$ are third order, but are scaled by an $\mathcal{O}_{\leq}(\kappa^2)$ term, so are $\mathcal{O}_{<}(\kappa^4)$. The next lowest order moments come from the product of the linear part of $r(x|z)$ with the cubic terms in $g(x,y|z)$, so are fourth order moments which are $\mathcal{O}_{\leq}(\kappa^4)$. All remaining terms are strictly higher order. Thus:
$$
\mathbb{E}_{X,Y}[r(X|z) g(X,Y|z)] = \mathcal{O}_{\leq}(\kappa^4)
$$
if $\nabla_x f(z,z) \neq 0$. If $\nabla_x f(z,z) = 0$ then $r(x|Z)$ does not have a linear term, so $\mathcal{O}_{<}(\kappa^4)$. In either case $\mathbb{E}_{X,Y}[r(X|z) g(X,Y|z)]$ converges to zero faster than $\mathbb{V}[r(X|z)]$ which we showed before was $\mathcal{O}_{=}(\kappa^2)$ when $\nabla_x f(z,z) \neq 0$ and is $\mathcal{O}_{=}(\kappa^4)$ when $\nabla_x f(z,z) = 0$. It follows that the higher order correction term in the denominator always converges to zero faster than the quadratic term, provided the quadratic term is not identically zero.  

Next, $\mathbb{E}_{X,Y}[(X - z)^{\intercal} H_{xy}(z,z) (Y - z) g(X,Y|z)]$ will involve, at lowest order, fifth order moments of the trait distribution, so is $\mathcal{O}_{\leq}(\kappa^5)$. Lastly, $\mathbb{E}_{X,Y}[g(X,Y|z)^2]$ will have lowest order central moments of order six, so is $\mathcal{O}_{\leq}(\kappa^5)$. Thus, both higher order correction terms in the numerator converge to zero faster than the quadratic term which we established was $\mathcal{O}_{=}(\kappa^4)$ provided $H_{xy} \neq 0$. 

Thus, both the numerator and denominator converge to their quadratic approximations unless $\nabla_x f(z,z) = 0$ and $H_{xx} = 0$, or $H_{xy} = 0$, proving the lemma statement. In the latter case, the quadratic approximation to $\epsilon$ equals zero, so, provided the numerator is nonzero, the true $\epsilon$ converges to its quadratic approximation (zero) at order $\mathcal{O}_{\leq}(\kappa)$. Otherwise the error terms in the numerator and denominator all converge to zero one order faster in $\kappa$ than the corresponding terms from the quadratic expansion, so the overall error from higher order terms vanishes one order faster in $\kappa$ than $\epsilon$, proving the lemma statement for $\epsilon$. 

The same arguments apply to $\mathbb{V}_{X,Y}[f(X,Y)]$. Expand $f(X,Y)$ into its quadratic approximation and the error contributed by higher order terms. Then the variance is an expectation of a square of a sum. Expanding the square, first term is the variance in the quadratic approximation. This is the term seen when making the quadratic approximation. The remaining terms are the errors in the quadratic approximation. The second term is the covariance between the quadratic approximation and the higher order terms. The last is the variance in the higher order terms. As the higher order terms are, at lowest order cubic, the resulting central moments are all higher order than the lowest order terms which determine the quadratic approximation. Under the chosen scalings these terms all vanish at least an order faster than quadratic approximation, so the error introduced by the higher order terms vanishes at least one order faster than the quadratic approximation. $\blacksquare$

\vspace{2 mm}
%%%%%%%%%%%%%%%%%%%%%%%%%%%%%%%%%%%%%%%%%%%
\textbf{Proof of Lemma 7: } Under the smoothness assumption there exists a ball centered at $z$ for any $z$ with nonzero radius $r(z)$ where the error in the local quadratic approximation is boundable using a power series whose lowest order terms are cubic. Under the smoothness assumptions, the support of the trait distribution is covered by a ball $B_{R(\kappa)}(z)$, centered at $z$ whose radius $R(\kappa) \rightarrow 0$ as $\kappa \rightarrow 0$. Thus there exists a small enough $\kappa$ such that $R(\kappa) < r(z)$. In that case, the support of the trait distribution is contained inside of the ball where $f(x,y)$ can be approximated quadratically, and the error in the quadratic approxmation is bounded. Thus we can replace $f(x,y)$ in all expectations with its quadratic approximation plus some error, then bound the error with a cubic function whose lowest order terms are cubic. If $f(x,y)$ admits a convergent Taylor series expansion on $B_{r(z)}(z)$, then the error bound could be built directly from the third and higher order terms in the Taylor series about $z$. We require second differentiability to ensure that the second order Taylor expansion (quadratic approximation) is well defined, but do not require higher order differentiability as we only need the errors in the quadratic approximation to vanish fast enough. 

Once $\kappa$ is small enough we replace $f(x,y)$ with its quadratic model plus higher order error terms inside each relevant expectation. Then we aim to show that the errors vanish faster than any of the expectations produced using the quadratic approximation. This is accomplished by showing that the bound on the error must vanish faster than any of the relevant expectations used in the quadratic approximation. The fastest vanishing terms in the quadratic approximations were $\mathcal{O}_{=}(\kappa^4)$ (numerator of $\epsilon$ or the denominator of $\epsilon$ when $\nabla_x f(z,z) = 0$). Therefore, to ensure that the higher order errors vanish fast enough we need to show that the resulting error terms converge, at slowest $\mathcal{O}_{<}(\kappa^4)$. 

Here we largely borrow the algebra we used for the previous lemma. First, replace the error terms with their power series bound. Then, that power series is, at lowest order cubic, so it behaves just like the higher order terms when $f(x,y)$ admitted a convergent Taylor expansion. Returning to our previous proof, we see that the resulting error terms all involved third, fourth, or fifth order moments, and vanished faster than their matching approximations provided third order moments vanished $\mathcal{O}_{<}(\kappa^2)$, fourth order moments vanished $\mathcal{O}_{<}(\kappa^3)$, and fifth order moments vanished $\mathcal{O}_{<}(\kappa^4)$. Then, as long as the higher order moments also vanished $\mathcal{O}_{<}(\kappa^4)$, every statement used to show convergence in the previous lemma applies, with a weaker rate guarantee. Before we could guarantee that all errors vanished an order faster than their terms from the quadratic approximation. Now we simply aim to show that they vanish faster than their matching quadratic approximation.

Thus, by leveraging the previous lemma proof, it is sufficient to show that, under our concentration assumptions, third order moments are $\mathcal{O}_{<}(\kappa^2)$, fourth order moments are $\mathcal{O}_{<}(\kappa^3)$, and fifth and higher order moments are $\mathcal{O}_{<}(\kappa^4)$. This scaling is guaranteed by the collapse of the support. 

Suppose that $R(\kappa) \rightarrow 0$ as $\kappa \rightarrow 0$. Note that $R(\kappa)$ must converge to zero slower than $\kappa$ since we has assumed that the trait covariance is $\mathcal{O}_{=}(\kappa^2)$, in which case there exist marginal trait distributions with standard deviation $\mathcal{O}_{=}(\kappa)$. It follows that the support of the distribution cannot vanish faster than $\mathcal{O}_{=}(\kappa)$. Let $\alpha, \beta$ be multi-indices and consider the absolute value of the central moment:
$$
\begin{aligned}
|\mathbb{E}_{X,Y}[(X-z)^{\alpha} (Y-z)^{\beta}]| & = \left|\int_{x \in \Omega} \int_{y \in \Omega} (x-z)^{\alpha} (y - z)^{\beta} \pi_x(x) \pi_x(y) dx dy \right| \\ & = \left| \int_{x \in B_{R(\kappa)}(z)} \int_{y \in B_{R(\kappa)}(z)} (x-z)^{\alpha} (y - z)^{\beta} \pi_x(x) \pi_x(y) dx dy \right| \\
& \leq \max_{x,y \in B_{R(\kappa)}(z)}\{|(x-z)^{\alpha}(y-z)^{\beta}|\} \leq R(\kappa)^{|\alpha| + |\beta|}
\end{aligned}
$$
where $w^{\alpha} = \prod_{j=1}^{T} w^{\alpha_j}$, $|\alpha| = \sum_j \alpha_j$, $|\beta| = \sum_j \beta_j$, and where $B_{R}(z)$ denotes the ball of radius $R$ centered at $z$. It follows that any central moment of order $|\alpha| + |\beta|$ must be $\mathcal{O}_{\leq}(R(\kappa)^{|\alpha| + |\beta|})$.

Our goal, then, is to ensure that $R(\kappa)$ goes to zero fast enough to ensure that $R(\kappa)^3 =\mathcal{O}_{<}(\kappa^2)$, $R(\kappa)^4 = \mathcal{O}_{<}(\kappa^3)$, and $R(\kappa)^n = \mathcal{O}_{<}(\kappa^4)$ for $n \geq 5$. The last requirement is the tightest, and is satisfied whenever $R(\kappa) = \mathcal{O}_{<}(\kappa^{4/5})$. If $R(\kappa) = \mathcal{O}_{<}(\kappa^{4/5})$ then $R(\kappa)^3 = \mathcal{O}_{<}(\kappa^{12/5}) = \mathcal{O}_{<}(\kappa^{2 + 2/5}) = \mathcal{O}_{<}(\kappa^2)$, $R(\kappa)^4 = \mathcal{O}_{<}(\kappa^{16/5}) = \mathcal{O}_{<}(\kappa^{3 + 1/5}) = \mathcal{O}_{<}(\kappa^2)$, and $R(\kappa)^n = \mathcal{O}_{<}(\kappa^{(4 n)/5}) = \mathcal{O}_{<}(\kappa^4)$ when $n \geq 5$. 

Thus, if $R(\kappa) = \mathcal{O}_{<}(\kappa^{4/5})$, then third order moments vanish faster than $\kappa^2$, fourth order moments vanish faster than $\kappa^3$, and fifth and higher order moments vanish faster than $\kappa^4$, ensuring that any errors left over by the quadratic approximation vanish faster than the approximation terms, thus proving the lemma. $\blacksquare$

\vspace{2mm}
%%%%%%%%%%%%%%%%%%%%%%%%%%%%%%%%%%%%%%%%%%%%%%%%%%%%

\textbf{Proof of Lemma 8: } Consider a bounded function $g(x,y)$ on $\Omega \times \Omega$. Then:
$$
\begin{aligned}
\mathbb{E}_{X,Y \sim \pi_x}[g(X,Y)] & = (1 - p(\kappa)) \mathbb{E}_{X,Y \sim \pi_x| X,Y \in B_{R(\kappa)}(z)}[g(X,Y)] + p(\kappa) \mathbb{E}_{X,Y \sim \pi_x| X,Y \notin B_{R(\kappa)}(z)}[g(X,Y)] \\
& = \mathbb{E}_{X,Y \sim \pi^w_x}[g(X,Y)] + p(\kappa) \left(\mathbb{E}_{X,Y \sim \pi_x| X,Y \notin B_{R(\kappa)}(z)}[g(X,Y)] - \mathbb{E}_{X,Y \sim \pi^w_x}[g(X,Y)] \right) .
\end{aligned}
$$

The first term is the windowed approximation to the expectation. The second term is the error in the windowed approximation. As long as $g(X,Y)$ is bounded, there exists a scalar $M$ such that $|g(X,Y)| \leq M$. Then any expectation of $g(X,Y)$ is also bounded by $\pm M$ so:
$$
|\mathbb{E}_{X,Y \sim \pi^w_x}[g(X,Y)] - \mathbb{E}_{X,Y \sim \pi_x}[g(X,Y)]| \leq 2 M p(\kappa).
$$

Therefore the error in the windowed approximation is $\mathcal{O}_{\leq}(p(\kappa)$. $\blacksquare$

\vspace{2mm}
%%%%%%%%%%%%%%%%%%%%%%%%%%%%%%%%%%%%%%%%%%%%%%%%%%%%%

\textbf{Proof of Theorem 2: } Apply Lemma 8 to Lemma 7, then substitute in the convergence rates of the expected components computed in Lemmas 4 and 5. $\blacksquare$

%% Bibliography
\bibliographystyle{siam}
\bibliography{Refs}

\end{document}